\documentclass[acmsmall,nonacm]{acmart}

\AtBeginDocument{%
  \providecommand\BibTeX{{%
    \normalfont B\kern-0.5em{\scshape i\kern-0.25em b}\kern-0.8em\TeX}}}

\setcopyright{acmcopyright}
\copyrightyear{2018}
\acmYear{2018}
\acmDOI{10.1145/1122445.1122456}

\acmJournal{JACM}
\acmVolume{37}
\acmNumber{4}
\acmArticle{111}
\acmMonth{8}

\usepackage{textcomp}
\newcommand{\slm}{\textrm{SLM}}
\usepackage{subcaption}
\usepackage{paralist}

\begin{document}

\title[Hate begets Hate]{Hate begets Hate: A Temporal Study of Hate Speech}


\author{Binny Mathew}
\email{binnymathew@iitkgp.ac.in}
\orcid{0000-0003-4853-0345}
\affiliation{%
  \institution{Indian Institute of Technology Kharagpur}
  \state{West Bengal}
  \country{India}
  \postcode{721302}
}

\author{Anurag Illendula}
\affiliation{%
  \institution{Indian Institute of Technology Kharagpur}
  \state{West Bengal}
  \country{India}
  \postcode{721302}
}
\email{aianurag09@gmail.com}

\author{Punyajoy Saha}
\affiliation{%
  \institution{Indian Institute of Technology Kharagpur}
  \state{West Bengal}
  \country{India}
  \postcode{721302}
}
\email{punyajoys@iitkgp.ac.in}

\author{Soumya Sarkar}
\affiliation{%
  \institution{Indian Institute of Technology Kharagpur}
  \state{West Bengal}
  \country{India}
  \postcode{721302}
}
\email{portkey1996@gmail.com}

\author{Pawan Goyal}
\affiliation{%
  \institution{Indian Institute of Technology Kharagpur}
  \state{West Bengal}
  \country{India}
  \postcode{721302}
}
\email{pawang@cse.iitkgp.ernet.in}

\author{Animesh Mukherjee}
\affiliation{%
  \institution{Indian Institute of Technology Kharagpur}
  \state{West Bengal}
  \country{India}
  \postcode{721302}
}
\email{animeshm@cse.iitkgp.ac.in}

\renewcommand{\shortauthors}{Mathew et al.}

\begin{abstract}
  With the ongoing debate on `freedom of speech' vs. `hate speech,' there is an urgent need to carefully understand the consequences of the inevitable culmination of the two, i.e., `freedom of hate speech' over time. An ideal scenario to understand this would be to observe the effects of hate speech in an (almost) unrestricted environment. Hence, we perform the first temporal analysis of hate speech on Gab.com, a social media site with very loose moderation policy. We first generate \textit{temporal snapshots} of Gab from millions of posts and users. Using these temporal snapshots, we compute an \textit{activity vector} based on DeGroot model to identify hateful users. The amount of hate speech in Gab is steadily increasing and the \textit{new users are becoming hateful at an increased and faster rate}. Further, our analysis analysis reveals that the hate users are occupying the prominent positions in the Gab network. Also, the language used by the community as a whole seem to correlate more with that of the hateful users as compared to the non-hateful ones. We discuss how, many crucial design questions in CSCW open up from our work.
\end{abstract}



\maketitle

\section{Introduction}

The question about where is the borderline or whether there is indeed any borderline between `free speech' and `hate speech' is an ongoing subject of debate which has recently gained a lot of attention. With crimes related to hate speech increasing in the recent times\footnote{\url{https://www.justice.gov/hatecrimes/hate-crime-statistics}}, it is considered to be one of the fundamental problems that plague the Internet. The online dissemination of hate speech has even lead to real-life tragic events such as the genocide of the Rohingya community in Myanmar, the anti-Muslim mob violence in Sri Lanka, and the Pittsburg shooting.
The big tech giants are also unable to control the massive dissemination of hate speech\footnote{\url{https://tinyurl.com/facebook-leaked-moderation}}.

Recently, there have been a lot of research concerning multiple aspects of hate speech such as \textit{detection}~\cite{davidson2017automated,Badjatiya:2017:DLH:3041021.3054223,zhang2018detecting}, \textit{analysis}~\cite{Chandrasekharan2017YouCS,Olteanu2018TheEO}, \textit{target identification}~\cite{silva2016analyzing,mondal2017measurement,elsherief2018hate}, \textit{counter-hate speech}~\cite{gagliardone2015countering,mathew2018thou,benesch2016countertwitter,mathew2020interaction} etc. However, very little is known about the temporal effects of hate speech in online social media, especially if it is considered as normative. In order to have a clear understanding on this, we would need to see the effects on a platform which allows free flow of hate speech. To understand the true nature of the hateful users, we need to study them in an environment that would not stop them from following/enacting on their beliefs. This led us to focus our study on Gab ($Gab.com$). Gab is a social media site that calls itself the \textit{`champion of free speech'}. The site has a loose moderation policy compared to other mainstream social media sites such as Twitter and Facebook, and does not prohibit a user from posting any hateful content. This naturally attracts users who want to express freely without moderation including hate speakers. This organic environment in which the main moderation is in the form what the community members impose on themselves provides a rich platform for our study.  Using a large dataset of $\sim 21M$ posts spanning around two years since the inception of the site, we develop a data pipeline which allows us to study the temporal effects of hate speech in a loosely moderated environment. Our work adds the temporal dimension to the existing literature on hate speech and tries to study and characterize hate in a loosely moderated online social media.

Despite the importance of understanding hate speech in the current socio-political environment, there is little CSCW work which looks into the temporal aspects of these issues. This paper fills an important research gap in understanding how hate speech evolves in an environment where it is protected under the umbrella of free speech. This paper also opens up questions on how new CSCW design policies of online platforms should be regulated to minimize/mitigate the problem of the temporal growth of hate speech. We posit that CSCW research, acknowledging the far-reaching consequences of this problem, should factor it into the ongoing popular initiative of \textit{platform governance}\footnote{\url{https://www.tandfonline.com/eprint/KxDwNEpqTY86MNpRDHE9/full}}.

\subsection{Outline of the work}
To understand the temporal characteristics, we needed data from consecutive time points in Gab. As a first step, using a heuristic, we generate successive graphs which capture the different time snapshots of Gab at one month intervals. Then, using the DeGroot model, we assign a hate intensity score to every user in the temporal snapshot and categorize them based on their degrees of hate. We then perform several \textit{linguistic} and \textit{network} studies on these users across the different time snapshots.

\subsection{Research questions}
\begin{compactenum}
\item \textbf{RQ1}: How can we characterize the growth of hate speech in Gab?
\item \textbf{RQ2}: How have the hate speakers affected the Gab community as a whole?
\end{compactenum}

RQ1 attempts to investigate the general growth of hate speech in Gab. Previous research on Gab~\cite{zannettou2018gab} states that the hateful content is 2.4x as compared to Twitter. RQ2, on the other hand, attempts to identify how these hateful users have affected the Gab community. We study this from two different perspectives: language and network characteristics.

\subsection{Key observations}
For RQ1, we found that the amount of hate speech in Gab is consistently increasing. This is true for the new users joining as well. We found that the recently joining new users take much less time to become hateful as compared to those that joined at earlier time periods. Further, the fraction of users becoming hateful is increasing as well.

For RQ2, we found that the language used by the community as a whole is becoming more correlated with that of the hateful users as compared to the non-hateful ones. The hateful users also seem to be playing a pivotal role from the network point of view.

\section{Prior Work}
The hate speech research has a substantial literature and it has recently gained a lot of attention from the computer science perspective. In the following sections, we will examine the various aspects of research on hate speech. Interested readers can follow Fortuna et al.~\cite{fortuna2018survey} and Schmidt et al.~\cite{ schmidt2017survey} for a comprehensive survey of this subject.

\subsection{Definition of hate speech}
Hate speech lies in a complex confluence of freedom of expression, individual, group and minority rights, as well as concepts of dignity, liberty and equality~\cite{gagliardone2015countering}. Owing to the subjective nature of this issue, deciding if a given piece of text contains hate speech is onerous. 
In this paper, we use the hate speech definition outlined in the work done by Elsherief et al.~\cite{elsherief2018hate}. The authors define hate speech as a ``direct and serious attack on any protected category of people based on their race, ethnicity, national origin, religion, sex, gender, sexual orientation, disability or disease.''. Others have a slightly different definition for hate speech but the spirit is roughly the same. 
In our work we shall mostly go by this definition unless otherwise explicitly mentioned.

\subsection{Related concepts}

Hate speech is a complex phenomenon, intrinsically associated to relationships among groups, and also relying on linguistic nuances~\cite{fortuna2018survey}. It is related to some of the concepts in social science such as incivility~\cite{maity2018opinion}, radicalization~\cite{agarwal2015using}, cyberbullying~\cite{chen2011detecting}, abusive language~\cite{chandrasekharan2017bag,nobata2016abusive}, toxicity~\cite{gunasekara2018review,srivastava2018identifying}, profanity~\cite{sood2012profanity} and extremism~\cite{mcnamee2010call}. Owing to the overlap between hate speech and these concepts, sometimes it becomes hard to differentiate between them~\cite{davidson2017automated}. Teh et al.~\cite{teh2018identifying} obtained a list of frequently used profane words from comments in YouTube videos and categorized them into 8 different types of hate speech. The authors aimed to use these profane words for automatic hate speech detection. Malmasi et al.~\cite{malmasi2018challenges} attempted to distinguish profanity from hate speech by building models with features such as $n$-grams, skip-grams and clustering-based word representations.

\if{0}
\subsection{Targets of hate speech}

Hate speech is intended to attack a person or group based on some protected attributes. Thus, it becomes necessary to understand these target groups more closely. Several studies have been conducted with the goal of describing online hate speech and which groups are more threatened. Silva et al. ~\cite{silva2016analyzing} performed a large scale study to understand the target of such hate speech on two social media platforms: Twitter and Whisper. The authors leverage the sentence structure to define a template of the form `I <intensity> <userintent> <hatetarget>' to identify hateful content in these social media with very high precision. After manually categorizing them into hate categories, the authors found that the top three hate categories in both the online platforms were the same - Race, behavior, and Physical. In an extension of this study, Mondal et al.~\cite{mondal2017measurement} performed a geographical analysis of hate speech in various countries as well. They found that 80\% of the hate speech in their Whisper dataset came from US alone. Among them there was a clear bias against black people. Kwok et al.~\cite{kwok2013locate} also studied racist hate speech against the black people on Twitter and developed a binary classifier.

 Elsherief et al.~\cite{elsherief2018hate} looked into directed and generalized hate speech and observed that directed hate speech is very personal, in contrast to Generalized hate speech, where religious and ethnic terms dominate. The generalized hate speech is dominated by hate towards religions as opposed to other categories, such as Nationality, Gender or Sexual Orientation. Elsherief et al.~\cite{elsherief2018peer} performed a comparative study of hate speech instigators and target users on Twitter. They curated a large dataset capturing various types of hate speech and studied distinctive characteristics of hate instigators and targets in terms of their profile self-presentation, activities, and online visibility. The authors found that hate instigators target more popular and high profile Twitter users, and that participating in hate speech can result in greater on-line visibility. A personality analysis of these two groups reveals that both groups have eccentric personality facets that differ from the general Twitter population. Sengun et al.~\cite{sengun2019analyzing} analyzed the hate speech expressed towards MENA players in Leagure of Legends computer game.

Another group of people affected by hate speech are the Refugees and Immigrants ~\cite{ross2017measuring, kreis2017refugeesnotwelcome, sanguinetti2018italian}. In Kreis et al.~\cite{kreis2017refugeesnotwelcome}, the authors analyzed the discursive strategies used in the tweets which include \emph{\#refugeesnotwelcome} on Twitter. They found that many of these tweets were argumentative in nature and used `Us vs Them' strategy~\cite{burnap2016us} to disseminate their hate speech. Sanguinetti et al.~\cite{sanguinetti2018italian} provide a corpus of 6,000 tweets annotated for hate speech against immigrants. Hate speech based on sex~\cite{hewitt2016problem,saha2018hateminers,ahluwalia2018detecting,bartlett2014misogyny} and gender~\cite{reddy2002perverts,gatehouse2018troubling} has also increased lately. Another major target of hate speech are people belonging to different religions. These majorly include people who are Jews~\cite{finkelstein2018quantitative,bilewicz2013harmful} and Muslims~\cite{awan2016islamophobia,vidgen2018detecting}. Users can become the target of hate speech based on Nationality~\cite{erjavec2012you} and Physical traits~\cite{sherry2016disability} as well.
\fi

\subsection{Effects of hate speech}

Previous studies have found that public expressions of hate speech affects the devaluation of minority members~\cite{greenberg1985effect}, the exclusion of minorities from the society~\cite{mullen2003ethnophaulisms}, psychological well-being and the suicide rate among minorities~\cite{mullen2004immigrant}, and the discriminatory distribution of public resources~\cite{fasoli2015labelling}. Frequent and repetitive exposure to hate speech has been shown to desensitize the individual to this form of speech and subsequently to lower evaluations of the victims and greater distancing, thus increasing outgroup prejudice~\cite{soral2018exposure}.

\subsection{Computational approaches}
The research interest in hate speech, from a computer science perspective, is gaining interest. Larger datasets~\cite{davidson2017automated,founta2018large,de2018hate} and different approaches have been devised by researchers to detect hateful social media comments. These methods include techniques such as dictionary-based~\cite{guermazi2007using},    distributional semantics~\cite{djuric2015hate}, multi-feature~\cite{salminen2018anatomy} and neural networks~\cite{Badjatiya:2017:DLH:3041021.3054223}.

Burnap et al.~\cite{burnap2016us} used a bag of words approach combined with hate lexicons to build machine learning classifiers. Gitari et al.~\cite{gitari2015lexicon} used sentiment analysis along with subjectivity detection to generate a set of words related to hate speech for its classification. Chau et. al~\cite{chau2007mining} used analysis of hyperlinks among web pages to identify hate group communities. Zhou et al.~\cite{zhou2005us} used  multidimensional scaling (MDS) algorithm  to  represent the  proximity  of hate  websites  and thus capture their level of similarity. Lie et al.~\cite{liu2015new} incorporated LDA topic modelling for improving the performance of the hate speech detection task. Saleem et al.~\cite{saleem2017web} proposed an approach to detecting hateful speech using self-identifying hate communities as training data for hate speech classifiers. 
Davidson et al.~\cite{davidson2017automated} used crowd-sourcing to label tweets into three categories: hate speech, only offensive language, and those with neither. 
Waseem et al.~\cite{waseem2016hateful} presented a list of criteria based on critical race theory to identify racist and sexist slurs.

More recently, researchers have started using deep learning methods~\cite{Badjatiya:2017:DLH:3041021.3054223,zhang2018detecting} and graph embedding techniques~\cite{ribeiro2018characterizing} to detect hate speech. 
Badjatiya et al.~\cite{Badjatiya:2017:DLH:3041021.3054223} applied several deep learning architectures and improved the benchmark score by $\sim$18 F1 points. 
Zhang et al.~\cite{zhang2018detecting} used deep neural network, combining convolutional and gated recurrent networks to improve the results on 6 out of 7 datasets. Gao et al.~\cite{gao2017detecting} utilized the context information accompanied with the text to develop hate speech detection models. Grondahl et al.~\cite{grondahl2018all} found that several of the existing state-of-the-art hate speech detection models work well only when tested on the same type of data they were trained on. Aluru et al.~\cite{aluru2020deep} perform a large scale analysis of multilingual hate speech detection in 9 languages from 16 different sources. The authors suggest that for low resource languages LASER + LR is more effective while for high resource BERT models are more effective.

While most of the computational approaches focus on detecting if a given text contains hate speech, very few works focus on the user account level detection. Gian et al.~\cite{qian2018leveraging} proposed a model that leverages intra-user and inter-user representation learning for hate speech detection. 
Gibson~\cite{gibson2017safe} studied the moderation policies on Reddit communities and observed that `safe space' have higher levels of censorship and is directly related to the politeness in the community. \citet{Seering:2019:BMP:3311957.3361855} explored moderation techniques to make the online communities more positive and supportive. \citet{hagen2019emoji} study the use of emojis in white nationalist conversation on Twitter and found striking difference between the `pro' and `anti' stance.

The work done by \citet{mathew2018spread} seem to be the closest to ours. We have utilized the same dataset, lexicon set, and parts of DeGroot model. However, \citet{mathew2018spread} studied the reach of hate speech on static following relationship graph. Our work utilizes a dynamic graph and investigates the temporal effects of hate speech. Our work uses temporal snapshots to find user position via core periphery analysis. Linguistic inclination of the community to hate speech is also found. 
Investigating the effects of hate speech in online social media remains an understudied area in CSCW research. By employing our data processing pipeline, we study the temporal effects of hate speech on Gab.

\section{Dataset}

\subsection{The Gab social network}

Gab is a social media platform launched in August 2016 known for promoting itself as the ``champion of free speech''. However, it has been criticized for being a shield for alt-right users~\cite{zannettou2018gab}. The site is very similar to Twitter but has very loose moderation policies. According to the Gab guidelines, the site does not restrain users from using hateful speech\footnote{\url{https://gab.com/about/tos}}. The site allows users to read and write posts of up to 3,000 characters. The site employs an upvoting and downvoting mechanism for posts and categorizes posts into different topics such as News, Sports, Politics, etc.

\subsection{Dataset collection}

We use the dataset developed by Mathew et al.~\cite{mathew2018spread} for our analysis. For the sake of completeness of the paper, we present the general statistics of the dataset in Table ~\ref{tab: dataset-details}. The dataset contains information from August 2016 to July 2018. We do not use the data for the initial two months (August-September 2016) and the last month (July 2018) as they had fewer posts.

\section{Methodology}

To address our research questions, we need to have a temporal overview of the activity of each user. So, our first task involves generating temporal snapshots to capture the month-wise activity of the users. We chose month-wise activity due to the dataset constraint\footnote{This constraint is present in other Gab datasets publicly released such as the pushshift dataset (\url{https://files.pushshift.io/gab}) and in \citet{fair2019shouting} } which provides the user creation date only in the `month-year' format. We chose `month' as it is the most fine grained unit available. This means that we cannot build the temporal graphs at a granularity finer than a month. While for the posts, creation date is available in much finer granularity (in seconds, minutes, and days), it does not suffice for the creation of the followership graph. 

We develop a pipeline to generate the \textit{temporal hate vectors} of each user for this purpose. A temporal hate vector is a representation used to capture the temporal activity of each user. A higher value in the hate vector is an indication of the hatefulness of a user, whereas a lower value indicates that the user potentially did not indulge in any hateful activity.

In this section, we will explain the pipeline we used to study the temporal properties of hate. The pipeline mainly consists of the following three tasks:

\begin{compactenum}
    \item \textbf{Generating temporal snapshots}: We divide the data such that a particular snapshot represents the activities of a particular month.
    \item \textbf{Hate intensity calculation}: We calculate the month-wise hate intensity score for each user, which represents the hateful activity of a user based on his/her posts, reposts, and network connections in a particular snapshot.
    \item \textbf{User profiling}: We profile users based on his/her temporal activity of hate speech, which is represented by a vector of his/her timeline of hate intensity scores.
\end{compactenum}

\begin{figure}
\centering
\begin{minipage}{\textwidth}

\begin{minipage}[b]{.60\textwidth}
  \centering
  \includegraphics[width=.95\linewidth]{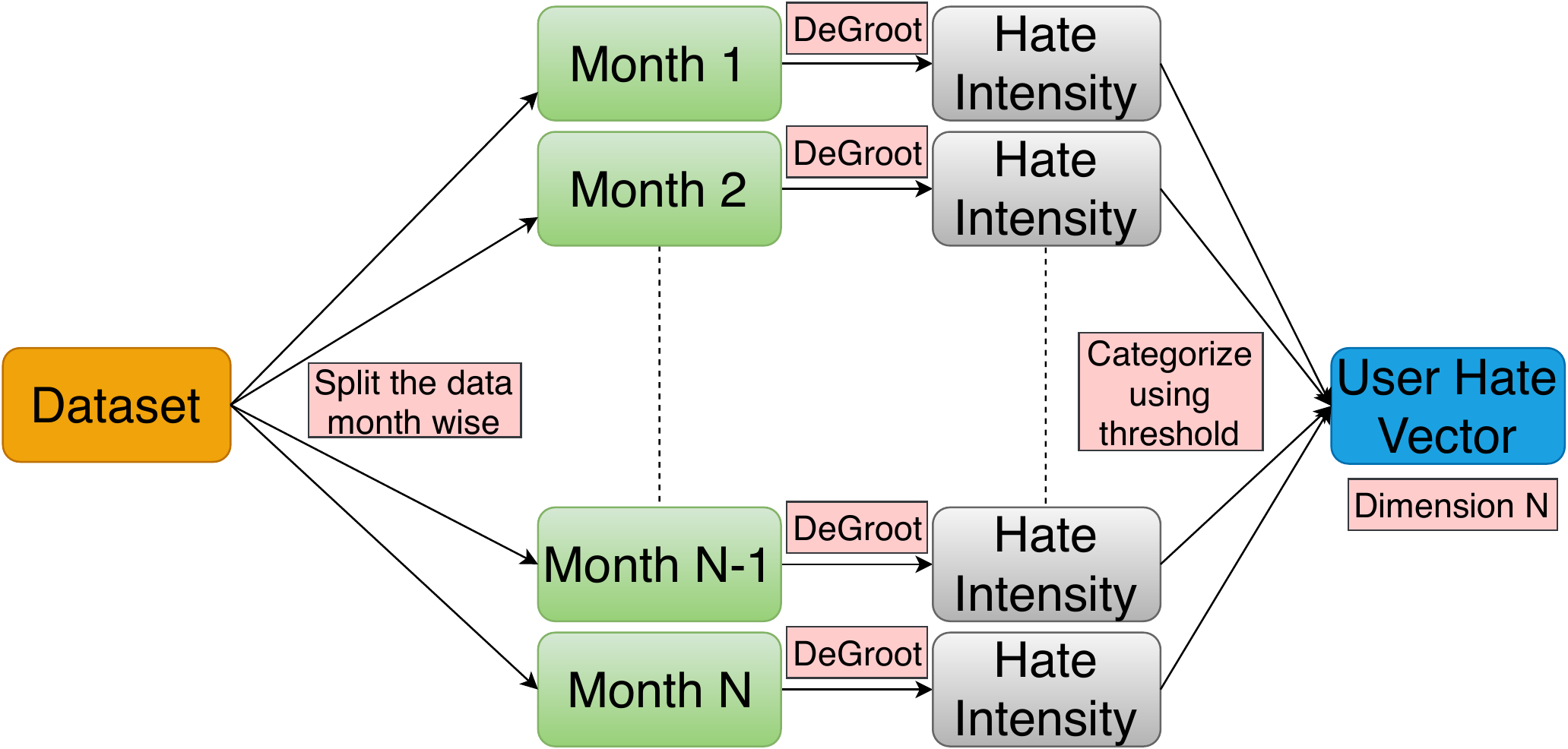}
  \captionof{figure}{Our overall methodology to generate the hate vector for a user. The dataset is first split month-wise before passing it to the DeGroot model. Using the DeGroot model, we calculate the month-wise `Hate Intensity' score for each user. These hate intensity scores are then used to generate the hate vector for each user.}
  \Description{Figure showing the dataset splitting into months and using the degroot model to calculate the hate intensity and generating user hate vector}
  \label{fig:DataProcessing}
\end{minipage}%
\hfill
\begin{minipage}[b]{.38\textwidth}
  \centering
\captionof{table}{Description of the dataset.}
    \resizebox{\linewidth}{!}{%
\begin{tabular}{l l}
\hline
Property & Value\\ \hline \hline
Number of posts&21,207,961\\
Number of reply posts&6,601,521\\
Number of quote posts&2,085,828\\
Number of reposts&5,850,331\\
Number of posts with attachments &9,669,374\\
Number of user accounts&341,332\\
Average follower per account & 62.56 \\
Average following per account & 60.93\\
\hline
\end{tabular}%
}
\label{tab: dataset-details}
\end{minipage}
\end{minipage}
\end{figure}

Figure \ref{fig:DataProcessing} shows our overall data processing pipeline.

\section{Generating Temporal Snapshots}
In order to study the temporal nature of hate speech, we need a temporal sequence of posts, reposts, users being followed, and users following the account. The Gab dataset hosts information regarding the post creation date, but it does not provide any information about when a particular user started following another user. Using various data points we have, we suggest a technique in the following section to approximate the month in which a user started following another user.

\subsection{New followers in each snapshot}

\begin{figure}
\centering
\begin{minipage}[t]{.45\textwidth}
  \centering
  \includegraphics[width=.9\linewidth]{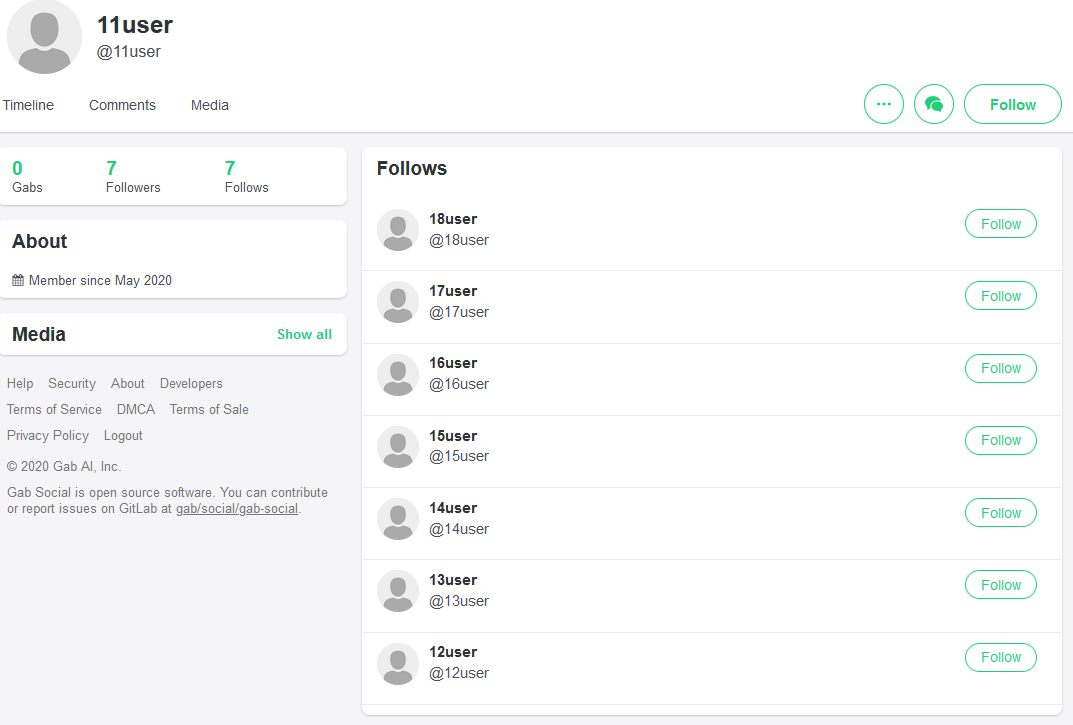}
  \captionof{figure}{The following list of 11user on Gab (\url{www.gab.com/11user/following}).}
  \Description{Figure showing the list of users that 11user is following}
  \label{fig:gabfollowsnapshot}
\end{minipage}%
\qquad
\begin{minipage}[t]{.45\textwidth}
  \centering
  \includegraphics[width=.9\linewidth]{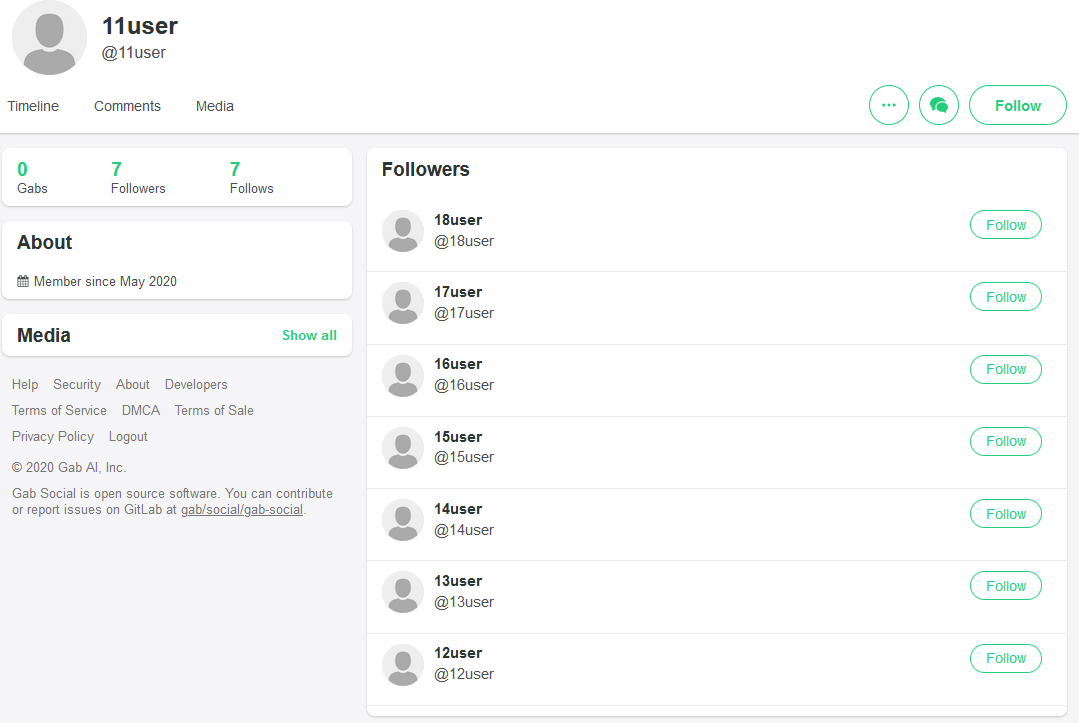}
  \captionof{figure}{The followers list of 11user on Gab (\url{www.gab.com/11user/followers}).}
  \Description{Figure showing the list of followers of 11user.}
  \label{fig:gabfollowersnapshot}
\end{minipage}
\end{figure}

While the post creation date is available in the dataset, the Gab API does not provide us with the information regarding when a particular user started following another user. The Gab API provides the followers/following information of a user in reverse chronological order. We have verified this manually by creating multiple accounts and following several Gab accounts. We created eight accounts on Gab with the usernames 1Xuser, where X ranges from 1 to 8. In Figure~\ref{fig:gabfollowsnapshot}, we made 11user account to follow other accounts in the order 12user, 13user, 14user, $\ldots$, 18user. As we can observe from the figure, Gab displays them in reverse chronological order (the latest followed account is on top of the list). Similarly, we made other users to follow 11user in the same order, i.e., 12user, 13user, 14user, $\ldots$, 18user. From Figure~\ref{fig:gabfollowersnapshot}, we can see that Gab provides the followers of 11user in reverse chronological order as well.

To get an estimate of when a user started following another user, we apply the heuristic developed by Meeder et al.~\cite{meeder2011we}, which was used in previous works~\cite{lang2011anti, antonakaki2015evolving} to get a lower bound on the \textit{following} link creation date. The heuristic is based on the fact that the API returns the list of followers/friends of a user ordered by the link creation time. We can thus obtain a lower bound on the follow time using the account creation date of a follower. For instance, if a user $U_A$ is followed by users $\{U_0, U_1, \ldots, U_n\}$ (in this order through time)\footnote{We have this information as the follower/following data is fetched in this manner.} and the users joined Gab on dates $\{D_0, D_1, \ldots, D_n\}$, then we can know for certain that $U_1$ was not following $U_A$ before $\max(D_0, D_1)$. We applied this heuristic on our dataset and ordered all of the \textit{following} relationships according to this. The authors~\cite{meeder2011we} proved that this heuristic is pretty accurate (within several minutes) specially on time periods where there are high follow rates. Since in our case we have considered a much larger window (one month), it would provide a fairly accurate estimate about the list of followers/friends each month for a particular user. This information, combined with the creation dates of her posts allows us to construct a temporal snapshot of his/her activity each month.

\subsection{Dynamic graph generation}\label{sec:dynamic graph}

\begin{figure}
\centering
\begin{minipage}[t]{.45\textwidth}
  \centering
  \includegraphics[width=.9\linewidth]{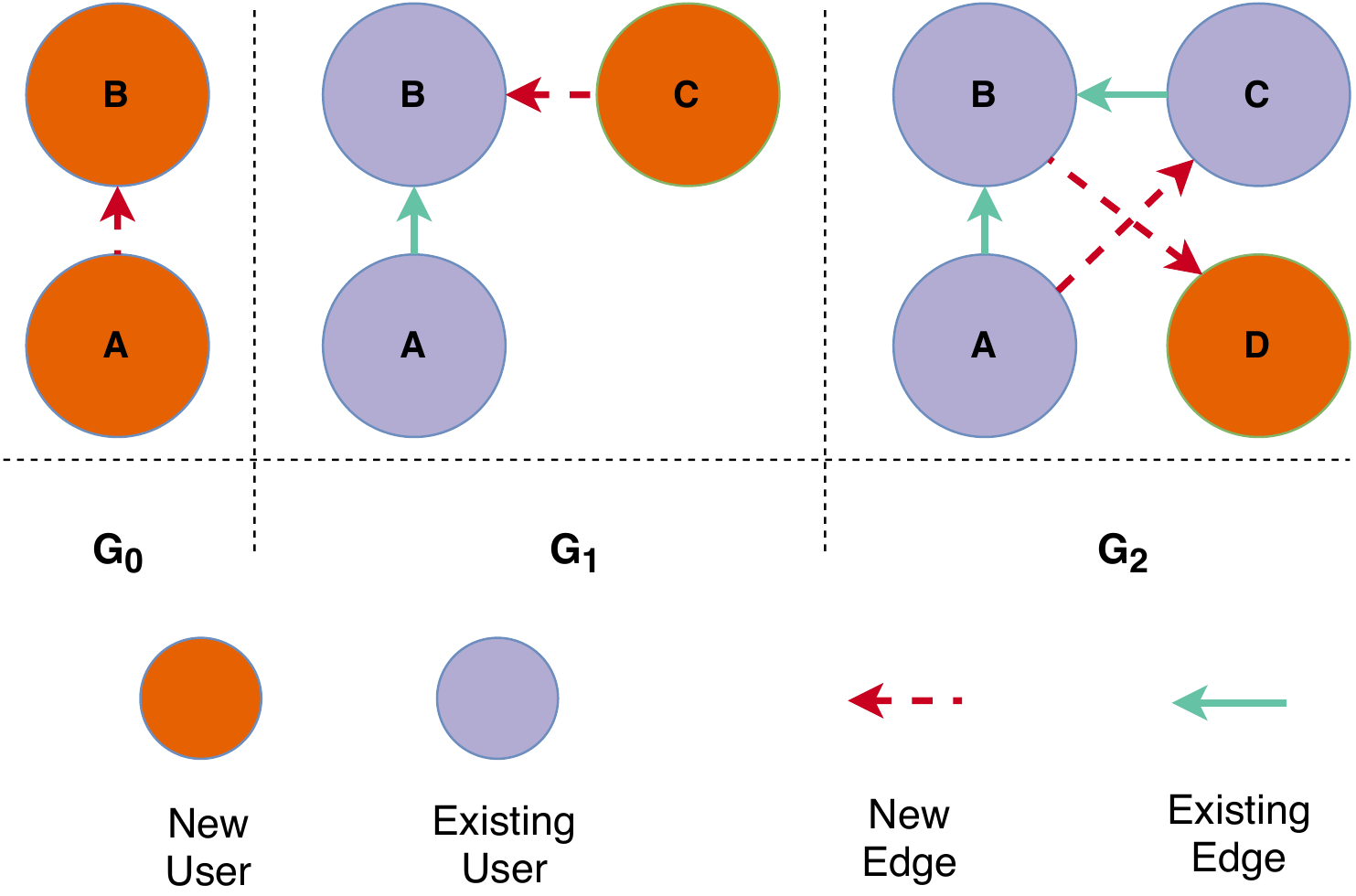}
  \captionof{figure}{An example dynamic graph. The nodes represent user accounts and the edges represent the `follows' relationship. Each successive snapshot is separated by one month duration.}
  \Description{Figure showing the temporal evolution of a network of four nodes.}
  \label{fig:DynamicGraph}
\end{minipage}%
\qquad
\begin{minipage}[t]{.45\textwidth}
  \centering
  \includegraphics[width=.9\linewidth]{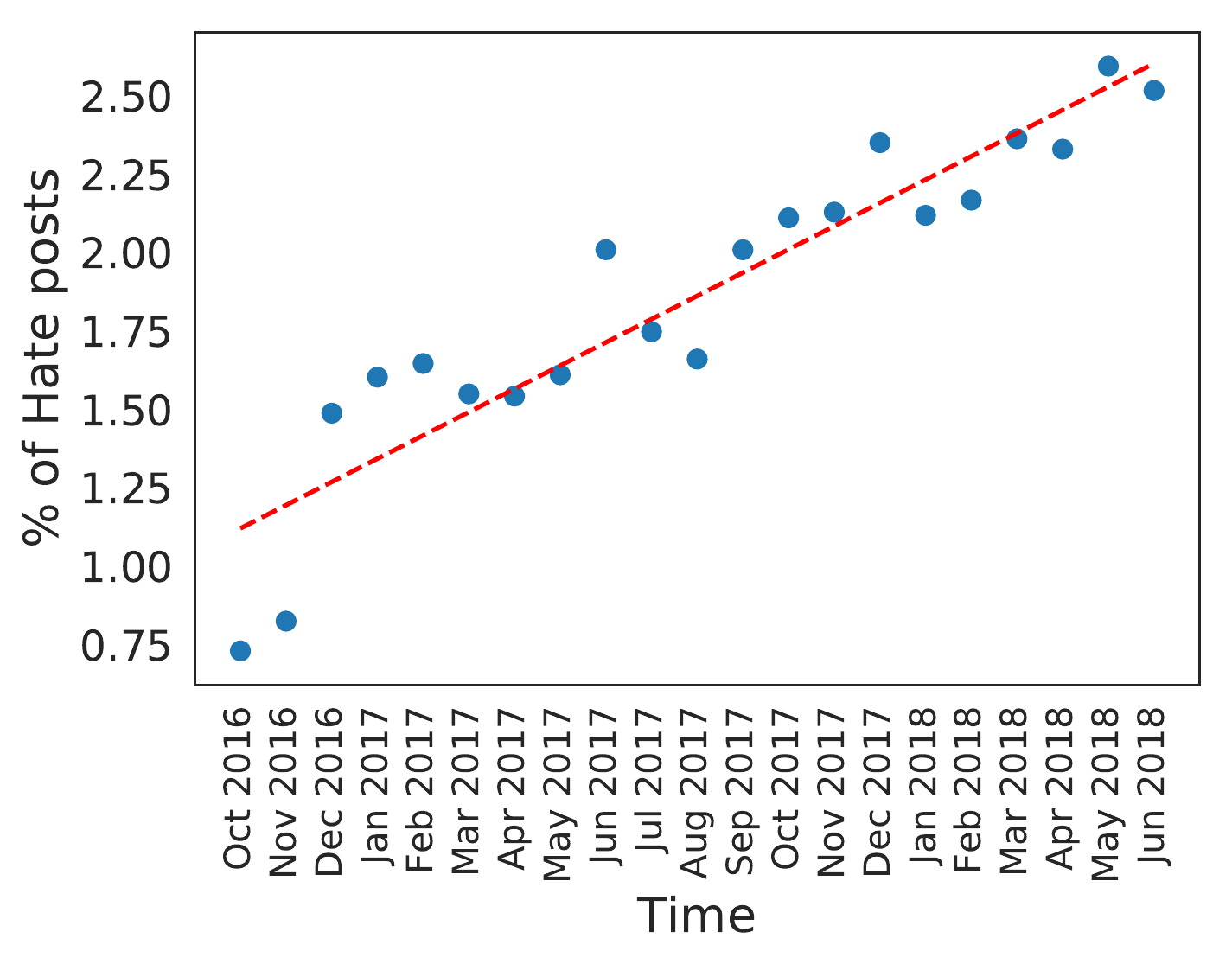}
  \captionof{figure}{The percentage of posts with atleast one hate word over monthly snapshots. The \textcolor{red}{red line} shows the increasing trend of posting such messages on Gab.}
  \Description{Plot with X axis as '\% of hate posts' and y axis as Time.}
  \label{fig:hate_post_increase}
\end{minipage}
\end{figure}

We consider the Gab graph ($\mathcal{G}$) as a dynamic graph with no parallel edges. We represent the \emph{dynamic graph} $\mathcal{G}$  as a set of successive time step graphs $\{G_0, \ldots, G_{t_\text{max}}\}$, where $G_s = (V_s, E_s)$  denotes the graph at snapshot $s$, where the set of nodes is $V_s$ (=$\{\bigcup_{i=0}^{s-1}V_i\}+\{\text{new nodes in snapshot } s\}$) and the set of edges is $E_s$ (=$\{\bigcup_{i=0}^{s-1}E_i\}+\{\text{new edges in snapshot } s\}$). An example of this dynamic graph is provided in Figure~\ref{fig:DynamicGraph}.

Each snapshot, $G_s$, is a weighted directed graph with the users as the nodes and the edges\footnote{ If $user_A$ unfollows $user_B$, then the edge has to be deleted. However, the graph does not consider the case of edge deletion since this information is not available in the data.} representing the following relationship. The edge weight is calculated based on the user's posting and reposting activity. We shall explain the exact mechanism of calculation of this weight in the following section.

\section{Hate Intensity calculation}
We make use of the temporal snapshots to calculate the \emph{hate intensity} of a user. The notion of \emph{hate intensity} allows us to capture the overall hatefulness of a user. A user with a high value of \emph{hate intensity} would be considered to be a potential hateful user as compared to another with lower value. The \textit{hate intensity} value ranges from 0 to 1, with 1 representing highly hateful user and values close to zero representing non-hateful user.

We use the DeGroot model~\cite{DeGroot1974reaching,golub2010naive,ribeiro2018characterizing,mathew2018spread} to calculate the \textit{hate intensity} of a user at each snapshot. Similar to Mathew et al.~\cite{mathew2018spread}, our purpose of using DeGroot model is to capture users who did not use the hate keywords explicitly, yet have a high potential to spread hate, by using their connections. We later perform manual evaluation to ensure the quality of the model. The DeGroot model requires users to have an initial belief value which is then updated using the model. In our case, we use a hate lexicon to assign these initial values.

\subsection{Hate lexicon} 

We initially started with the lexicon set available in Mathew et al.~\cite{mathew2018spread}. These high-precision keywords were selected from Hatebase\footnote{\url{https://hatebase.org}} and Urban dictionary\footnote{\url{https://www.urbandictionary.com}}.
To further enhance the quality of the lexicon, we adopt the word embedding method, skip-gram~\cite{mikolov2013distributed}, to learn distributed representation of the words from our Gab dataset in an unsupervised manner. This would allow us to enhance the hate lexicon with words that are specific to the dataset as well as spelling variations used by the Gab users. For example, we found more than five variants for the derogatory term \emph{ ni**er} in the dataset used by hateful users. We manually went through the words and carefully selected only those words which could be used in a hateful context. This resulted in a final set of 187 phrases which we have made public\footnote{\url{https://github.com/binny-mathew/HateBegetsHate_CSCW2020}}
for the use of future researchers. 

\noindent\textbf{Quality of the hate lexicon}: In order to establish the quality of this lexicon, we adopt a stratified approach by collecting randomly 0.5\% of the posts for each keyword based on its frequency. This will ensure that high-frequency words are sampled and labeled more. Two of the authors independently annotated these posts and yielded an agreement of 88.5\%.  In agreed upon samples, 95\% were hateful. These values indicate that the lexicons developed are of high quality. The annotators were instructed to follow the definition of hate speech used in Elsherief et al.~\cite{elsherief2018hate}. Further our lexicon is time agnostic. To validate we select 50 high belief score users from first and last month, take 1\% posts and label them as hate/non-hate. Difference in recall of our keywords in identifying hate posts at the two time points is only .01.

In Figure~\ref{fig:hate_post_increase}, we plot the \% of posts that have at least one of the words from this hate lexicon. We can observe from these initial results that the volume of hateful posts on Gab is increasing over time.

\subsection{DeGroot model}

In the DeGroot opinion dynamics model~\cite{DeGroot1974reaching}, each individual has a fixed set of neighbours, and the local interaction is  captured  by  taking the  convex combination  of  his/her  own  opinion  and  the  opinions  of  his/her neighbours at  each  time  step~\cite{xu2015modified}. The DeGroot model describes how each user repeatedly  updates  her  opinion  to  the  average of  those  of  its neighbours.  Since  this  model  reflects  the  fundamental  human cognitive capability of taking convex combinations when integrating related information~\cite{anderson1981foundations}, it has been studied extensively in the past decades~\cite{chamley2013models}. We will now briefly explain the DeGroot model and how we adapt it to calculate the \emph{hate intensity} of a user account.

In the DeGroot model\footnote{All our network snapshots contain strongly connected subgraphs and are aperiodic thus justifying use of DeGroot model.}, each user starts with an initial belief. In each time step, the user interacts with its neighbours and updates his/her belief based on the neighbour's beliefs. Recall that each snapshot is a directed graph, $G_s=(V_s,E_s)$ with $V_s$ representing the set of vertices and $E_s$ representing the set of edges at snapshot $s$. Let $N(i)$ denote the set of neighbours of node $i$ and $z_i(t)$ denote the belief of the node $i$ at iteration $t$. The initial value of $z_i(t)$ is assigned based on the hate lexicons. The update rule in this model is the following: $\mathbf{z}_{i}(t+1)= \frac{w_{ii}\mathbf{z}_{i}(t) + \sum_{j\in N(i)}w_{ij}\mathbf{z}_{j}(t)} {w_{ii}+\sum_{j\in N(i)}w_{ij}}$ where $(i,j)\in E_s$.

For each snapshot, we assign the initial edge weights based on the following criteria:

\begin{align}\small\label{weight_update}
    w_{ij} =
  \begin{cases}
     \mathrm{e}^\text{ $R_{ij}$ }       & \quad \text{if $F_{ij}=1$}\\
    \mathrm{e}^\text{ $R_{ij}$ }       & \quad \text{if $F_{ij}=0$ and $R_{ij}>0$}\\
    0       & \quad \text{if $F_{ij}=0$ and $R_{ij}=0$}\\
    1+ \text{ $P_i$}  & \quad \text{if $i$ = $j$}\\
  \end{cases}
\end{align}

where $R_{ij}$ denotes the number of reposts done by user $i$, where the original post was made by user $j$. $F_{ij}$ represents the following relationship, where $F_{ij}=1$ means that user $i$ is following user $j$, and $F_{ij}=0$ means that user $i$ is not following user $j$.  Similarly, $P_i$ denotes the number of posts by user $i$.
If a user is inactive (no posts or reposts), then we put the value of $R_{ij}$ as zero in Equation 1. Thus, an inactive user will have a value which is an average of its neighbourhood value. If an inactive user is connected to lots of hateful users, its hate score will be high, even if it is inactive. We have kept it as such because the inactive here simply means no posts. However, the user can still follow users and like their posts.

We then run the DeGroot model on each snapshot graph for 5 iterations, similar to Mathew et al.~\cite{mathew2018spread}, to obtain the hate score for each of the users. Our network is based on repost mechanism. User A (following user B) will repost user B, if it agrees with the stance of A. We exponentiated this by assigning number of reposts as edge weight. Thus if A’s value aligns with B, it will repost more of B’s content as in other social networks. This is exactly what the DeGroot model attempts to enforce thus justifying the choice of this model.

\subsection{Calculating the hate score}
Using the high precision hate lexicon directly to assign a hate score to a user should be problematic because of two reasons: first, we might miss out on a large set of users who might not use any of the words in the hate lexicon directly or use spelling variations, thereby, getting a much lower score. Second, many of the users share hateful messages via images, videos and external links. Using the hate lexicon for these users will not work. Instead, we use a variant of the methodology used in Riberio et al.~\cite{ribeiro2018characterizing} to assign each user in each snapshot a value in the range $[0, 1]$ which indicates the users' propensity to be hateful.

We enumerate the steps of our methodology below. We apply this procedure for each snapshot to get the hate score for each user.

\begin{compactitem}
\item We identify the initial set of potential hateful users as those who have used the words from the hate lexicon in at least two posts.
Rest of the users are identified as non-hateful users.

\item Using the snapshot graph, we assign the edge weight according to Equation~\ref{weight_update}. We convert this graph into a belief graph by reversing the edges in the original graph and normalizing the edge weights between $0$ and $1$.

\item We then run a diffusion process based on the DeGroot's learning model on the belief network. We assign an initial belief value of $1$ to the set of potential hateful users identified earlier and $0$ to all the other users.

\item We observe the belief values of all the users in the network after \emph{five} iterations of the diffusion process. 

\end{compactitem}

\subsection{Threshold selection}
The DeGroot's model assigns each user a hate score in the range $[0,1]$ with $0$ implying the least hateful and $1$ implying highly hateful. In order to draw the boundary between the hateful and non-hateful users, we need a threshold value, above which we might be able to call a user as hateful. The same argument goes for the non-hateful users as well: a threshold value below which the user can be considered to be non-hateful.

\begin{figure}[!t]
\centering
\begin{minipage}[t]{.45\textwidth}
  \centering
  \includegraphics[width=.9\linewidth]{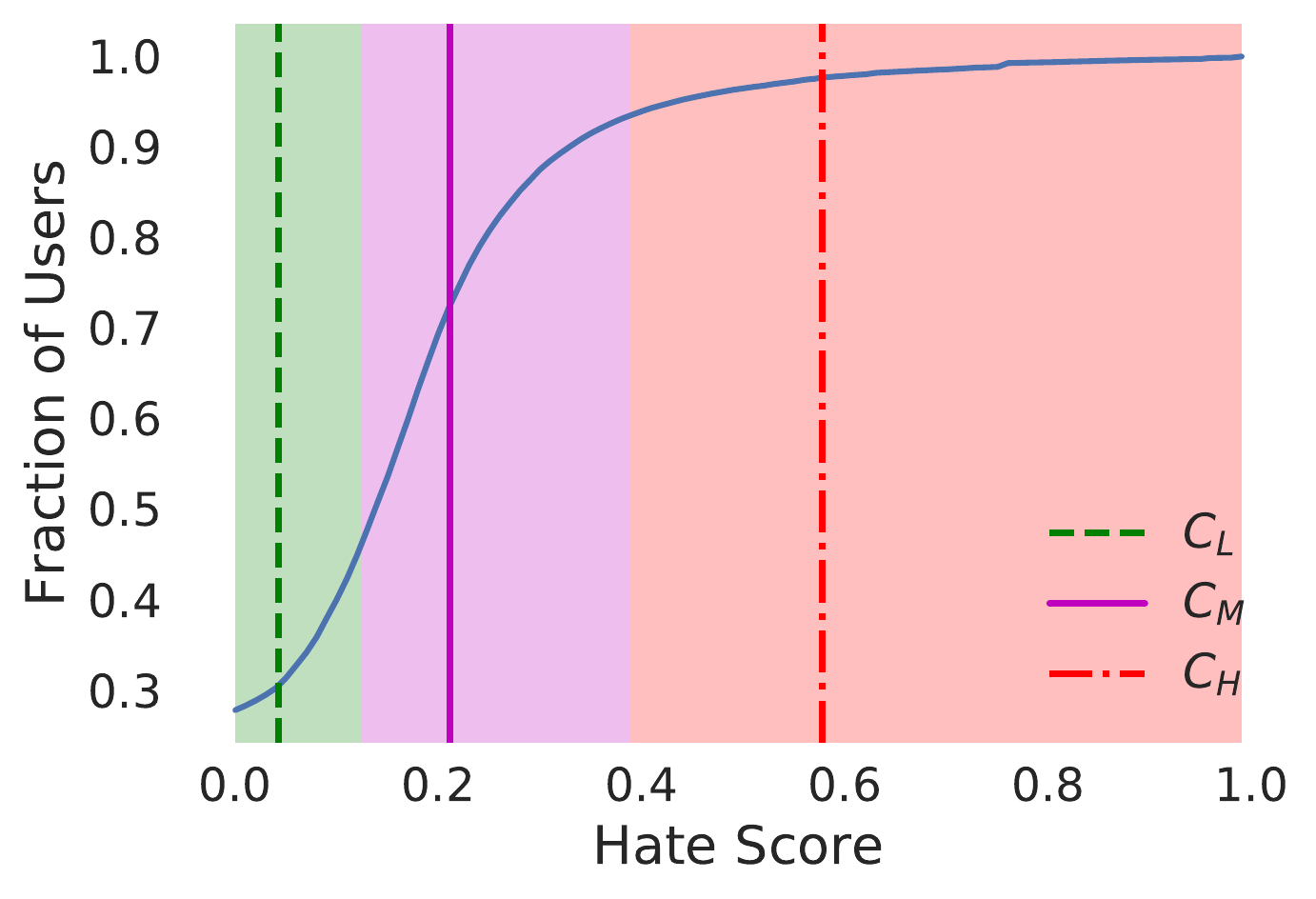}
  \captionof{figure}{The cumulative distribution of hate scores and the centroid values based on the $k$-means algorithm.  $C_L (=0.0421)$, $C_M (=0.2111)$, and $C_H (=0.5778)$ represent the centroids for the Low, Mid and High scores, respectively.}
  \Description{Figure with X-axis "Fraction of Users" and Y-axis "Hate Score". }
  \label{fig:hate_score_kmeans}
\end{minipage}%
\quad
\begin{minipage}[t]{.45\textwidth}
  \centering
  \includegraphics[width=.9\linewidth]{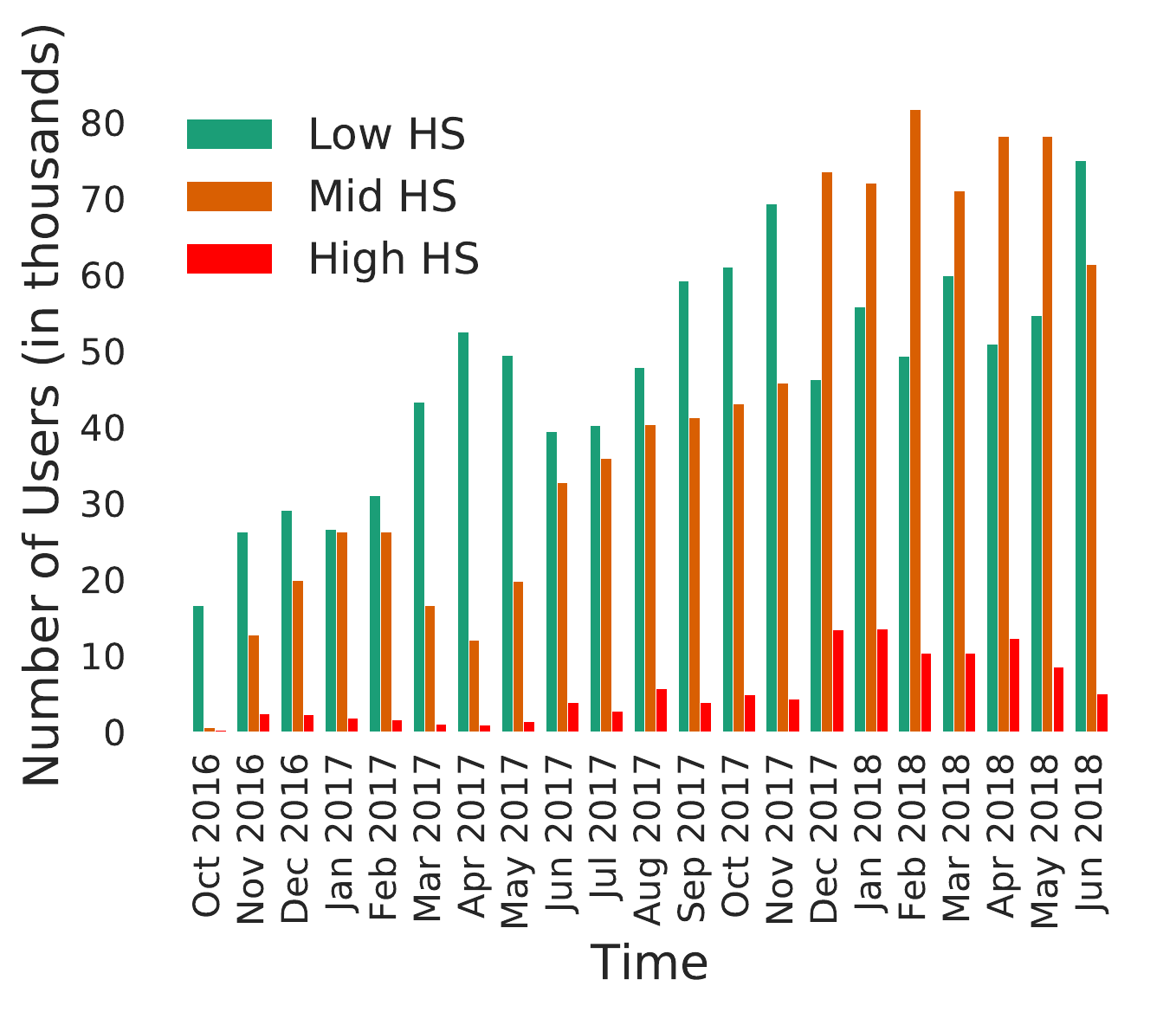}
  \captionof{figure}{The number of accounts that are labelled as low, mid, and high hate in each of the snapshots.}
  \Description{Plot with X-axis "Number of Users" and Y-axis "Time".}
  \label{fig:hate_distribution}
\end{minipage}%
\end{figure}

In order to select such threshold values, we used $k$-means~\cite{macqueen1967some,jain1999data} as a clustering algorithm on the scalar values of the hate score. Briefly, $k$-means selects $k$ points in space to be the initial guess of the $k$ centroids. Remaining points are then allocated to the nearest centroid. The whole procedure is repeated until no points switch cluster assignment or a number of iterations is performed.
In our case, we assign $k=3$ which would give us three regions in the range $[0,1]$ represented by three centroids $C_L$, $C_M$, and $C_H$ denoting `low hate', `medium hate' and `high hate', respectively. The purpose of having medium hate category is to capture the ambiguous users. These will be the users who will have values that are neither high enough to be considered hateful nor low enough to be considered non-hateful. We apply $k$-means algorithm on the list of hate scores from all the snapshots. Figure~\ref{fig:hate_distribution} shows the fraction of users in each category of hate in each snapshot (see the Discussion section for in-depth analysis of this figure). The DeGroot model is biased toward non-hate users as in every snapshot, a substantial fraction of users are initially assigned a value of zero. As shown in Figure~\ref{fig:hate_score_kmeans}, the centroid values are 0.0421 ($C_L$), 0.2111 ($C_M$), 0.5778 ($C_H$) for the low, mid, and high hate score users, respectively.

\section{User profiling}

Using the centroid values ($C_L$, $C_M$, and $C_H$), we transform the activities of a user into a sequence of low, medium, and high hate over time.
We denote this sequence by a vector $V_{hate}$. Each entry in $V_{hate}$ consists of one of the three values of low, mid, and high hate. This would allow us to find the changes in the perspective of a user at multiple time points.

\begin{figure}[bt]   	
	\centering
    
	\begin{subfigure}[t]{\linewidth}
        \includegraphics[width=\textwidth]{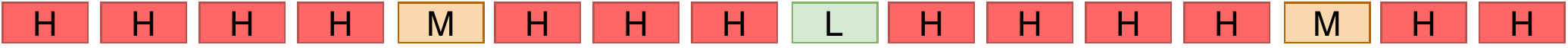}
	    \caption{Hateful user}
	    \Description{Figure showing the vector of hateful users. Majority of the values are high hate.}
	    \label{fig:hateful_user_vector}
	\end{subfigure}

	\begin{subfigure}[t]{\linewidth}
        \includegraphics[width=\textwidth]{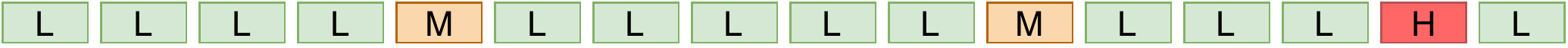}
	    \caption{non-hateful user}
	    \Description{Figure showing the vector of non-hateful users. Majority of the values are low hate.}
	    \label{fig:non_hateful_user_vector}
	\end{subfigure}
	
\caption{The hate vector consisting of sequence of low (L), mid (M), and high (H) hate. (a) An example of a hateful user as at least 75\% of the entries are High Hate (H) in the hate vector (b) An example of a non-hateful user as at least 75\% of the entries are Low Hate (L)} 
  \label{fig:hate_sequence_example}
\end{figure}

Consider the example given in Figure~\ref{fig:hateful_user_vector}. The vector represents a user who had high hate score for most of the time period with intermittent low and medium hate scores. Similarly, Figure~\ref{fig:non_hateful_user_vector} shows a user who had low hate score for most of the time period. 
For the purpose of this study, we mainly focus on only two types of user profiles: the consistently hateful users and the consistently non-hateful users.

We would like to point out here that other types of variations could also be possible. Like a user's hate score might change from one category to other multiples times, but we have not considered such cases here.

In order to find these users, we adopt a conservative approach and categorize the users based on the following criteria:

\noindent\textbf{Hateful}: We would call a user as hateful if at least 75\%\footnote{We verified that changing this threshold to 70\% or 80\% does not affect the qualitative observations. We thus continue with this threshold for high precision.} of his/her $V_{hate}$ entries contain an `H'.

\noindent\textbf{Non-hateful}: We would call a user as non-hateful user if at least 75\% of his/her $V_{hate}$ entries contain an `L'.

In addition, we used the following filters on the user accounts as well:
\begin{compactitem}
    \item The user should have posted at least \textbf{five} times.
    \item The account should be created before February 2018 so that there are at least \textbf{six} snapshots available for the user.
\end{compactitem}

We have not considered users with hate score in the mid-region as they are ambiguous. 
After the filtering, we got 1,019 users as hateful and 19,814 users as non-hateful.
In the following section, we perform textual and network analysis on these types of users and try to characterize them.

\subsection{Descriptive statistics}
We present various statistics of the dataset created and compare the hateful and the non-hateful users.

\noindent\textbf{Average number of posts}: The average number of posts made by hateful users per month is 4.32 times more than non-hateful users.

\noindent\textbf{Replies per user}: Out of the total posts with replies, hateful users have received 212 replies per user and non-hateful users have 47 replies per user.

\noindent\textbf{Topics of post}: To identify the topics of interest of the hateful and the non-hateful users, we used an LDA model to induce topics. We manually analysed the top 10 topics from LDA in text of non-hateful and hateful users. We found that hateful users tend to speak about Blacks, Muslims, Jews and politics. Though the non-hateful users also speak about politics, their topics include technology, sports and free speech on Gab.

In summary we observed that the hateful users are very different from the non-hateful users. These differences pave the way to the two research questions that we put forward next.

\subsection{Sampling the appropriate set of non-hateful users}

We use the non-hateful users as the candidates in the control group. Our idea of the control group is to find non-hateful users with similar characteristics as the hateful users. For sanity check purpose, we identify users who have (nearly) the same activity rate as the users in the hate set. We define the activity rate of a user as the sum of the number of posts and reposts done by the user, divided by the account age as of June 2018. For each hateful user, we identify a user from the non-hateful set with the nearest activity rate. We repeat this process for all the users in the hate list. We then perform Mann-Whitney $U$-test~\cite{depuy2005w} to measure the goodness. We found the value of $U = 517,208$ and $p$-value = 0.441. This indicates that the hateful and non-hateful users have nearly the same activity distribution. By using this subset of non-hateful users, we aim to capture any general trend in Gab. Our final set consists of 1,019 hateful users and the corresponding 1,019 non-hateful users who have very similar activity profile. We observe that these 2,038 (0.4\% of total) users produce 14.1\% posts on Gab attesting the power law effect (similar to \citet{mathew2018spread}). So, we have taken a highly active set in the Gab community for our study. But we would like to clarify that we are not extrapolating this to the entire Gab community as there are multiple topics of interest in Gab and hate is just one among these. However, as it stands now, hate makes a substantial part in this community and thus qualifies as an independent subject of study.

\subsection{Evaluation of user profiles}

We evaluate the quality of the final dataset of hateful and non-hateful accounts through human judgment. We ask two annotators to determine if a given account is hateful or non-hateful as per their perception. 

Since Gab does not have any policy for hate speech, we use the definition of hate speech provided by Elsherief et al.~\cite{elsherief2018hate}
for this task. Out of 1,019 hateful and 19,814 non-hateful users obtained from our approach, we provide the annotators with a class balanced random sample of 200 user accounts for each of the two classes.

Each of these 400 accounts was evaluated by two independent annotators and in the case of a tie, a third annotator was used (in 142 cases). For these 142 cases, we had to get an expert with deep knowledge in hate speech science to adjudicate and break the ties. We compared these manual annotations with our model outcomes and found the accuracy of 73\% for detecting hateful accounts and 67\% for detecting non-hate accounts.

\section{RQ1: How can we characterize the growth of hate speech in Gab?}

\subsection{The volume of hate speech shows an increasing trend}
As a first step to measure the growth of hate speech in Gab, we use the hate lexicon that we generated to find the number of posts which contain them in each month. We can observe from Figure~\ref{fig:hate_post_increase} that the amount of hate speech in Gab is indeed increasing.

\subsection{Higher fraction of new users are becoming hateful}
\begin{figure}[!t]

   \includegraphics[width=0.5\linewidth]{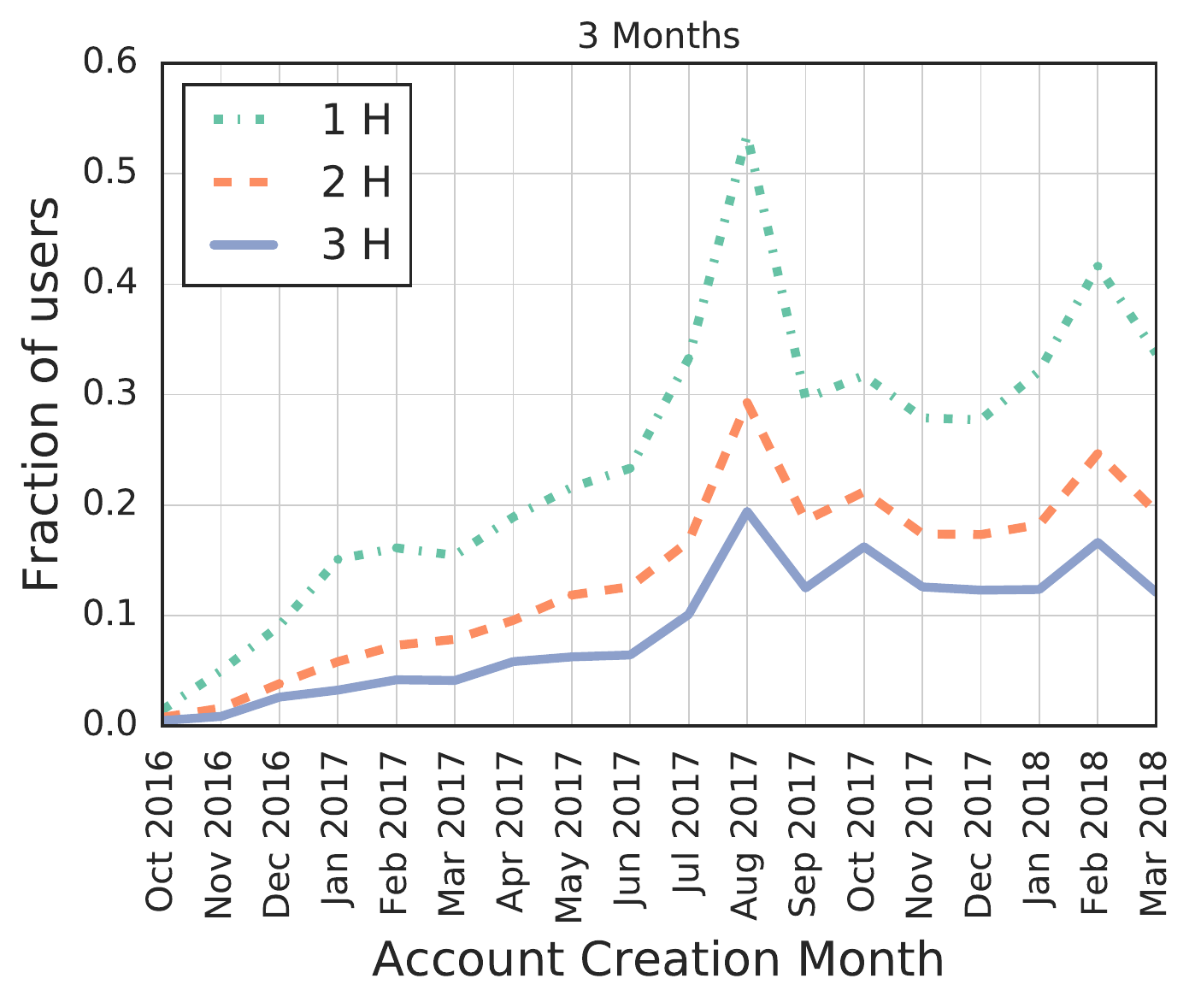} 
   \caption{The figure shows the fraction of users in each month who got assigned at least \textbf{(1, 2, 3)} `H' within the first \textbf{three} months of their joining. We observe that with time, a higher fraction of new users are using hateful language.}
   \Description{Figure with X-axis "Fraction of user" and Y-axis "Account Creation Month".}
   \label{fig:hate_score_from_creation_month}
\end{figure}

Another important aspect about the growth that we considered was the fraction of new users who are becoming hateful. In this scenario, we say that a user $A$ has become hateful, if his/her hate vector has the entry `H' at least $\mathcal{N}$ times within $\mathcal{T}$ months from the account creation. In Figure~\ref{fig:hate_score_from_creation_month}, we plot for $\mathcal{T}=3$ and $\mathcal{N}=1, 2, 3$ `H' entries, to observe the fraction of users for each month who are becoming hateful. As we can observe, the fraction of new users using hate speech is increasing over time.

\begin{figure*}[t]   	
	\centering
    
	\begin{subfigure}[t]{0.45\textwidth}
        \includegraphics[width=\textwidth]{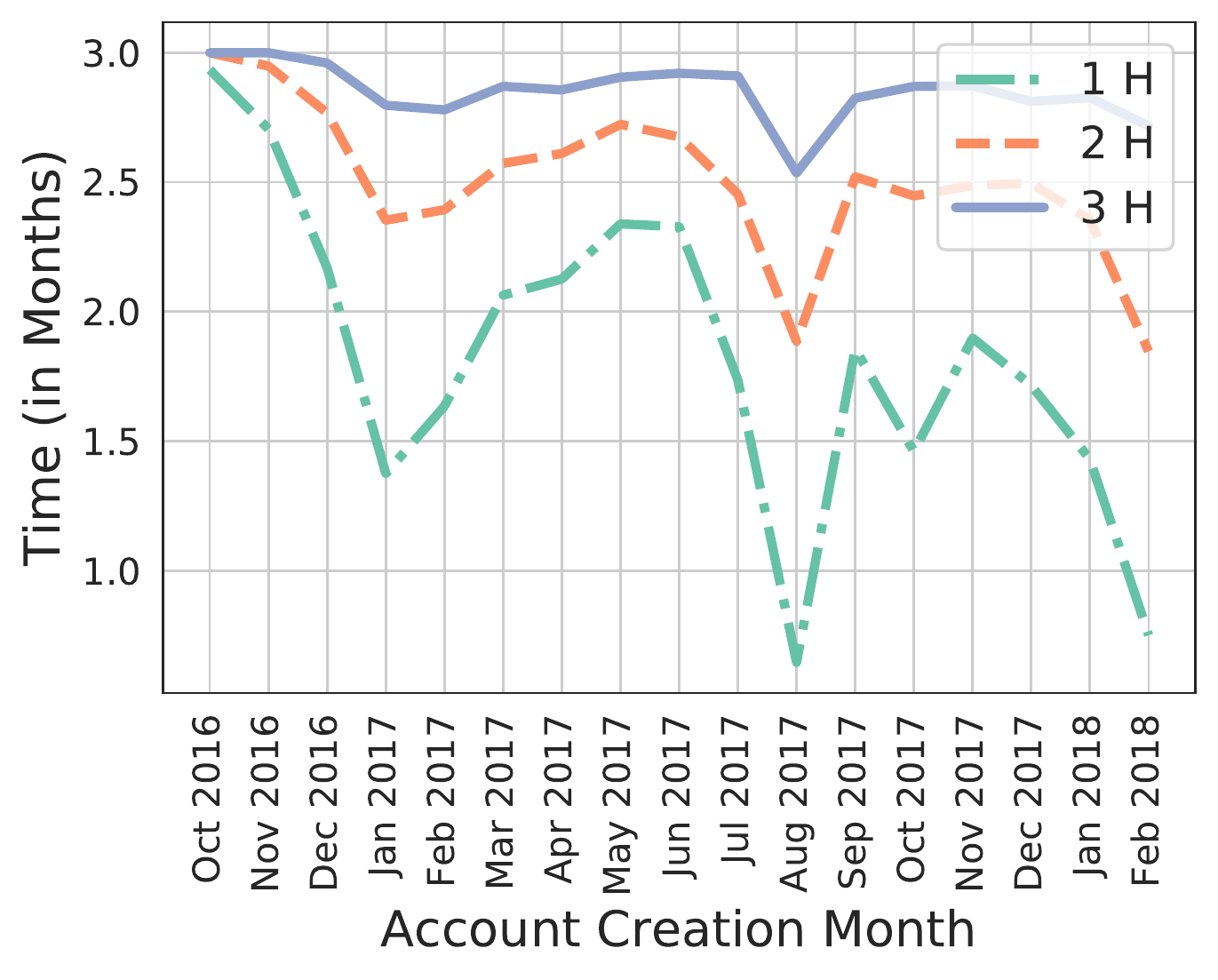}
	    \caption{Threshold = 3}
	\end{subfigure}
\begin{subfigure}[t]{0.45\textwidth}	        \includegraphics[width=\textwidth]{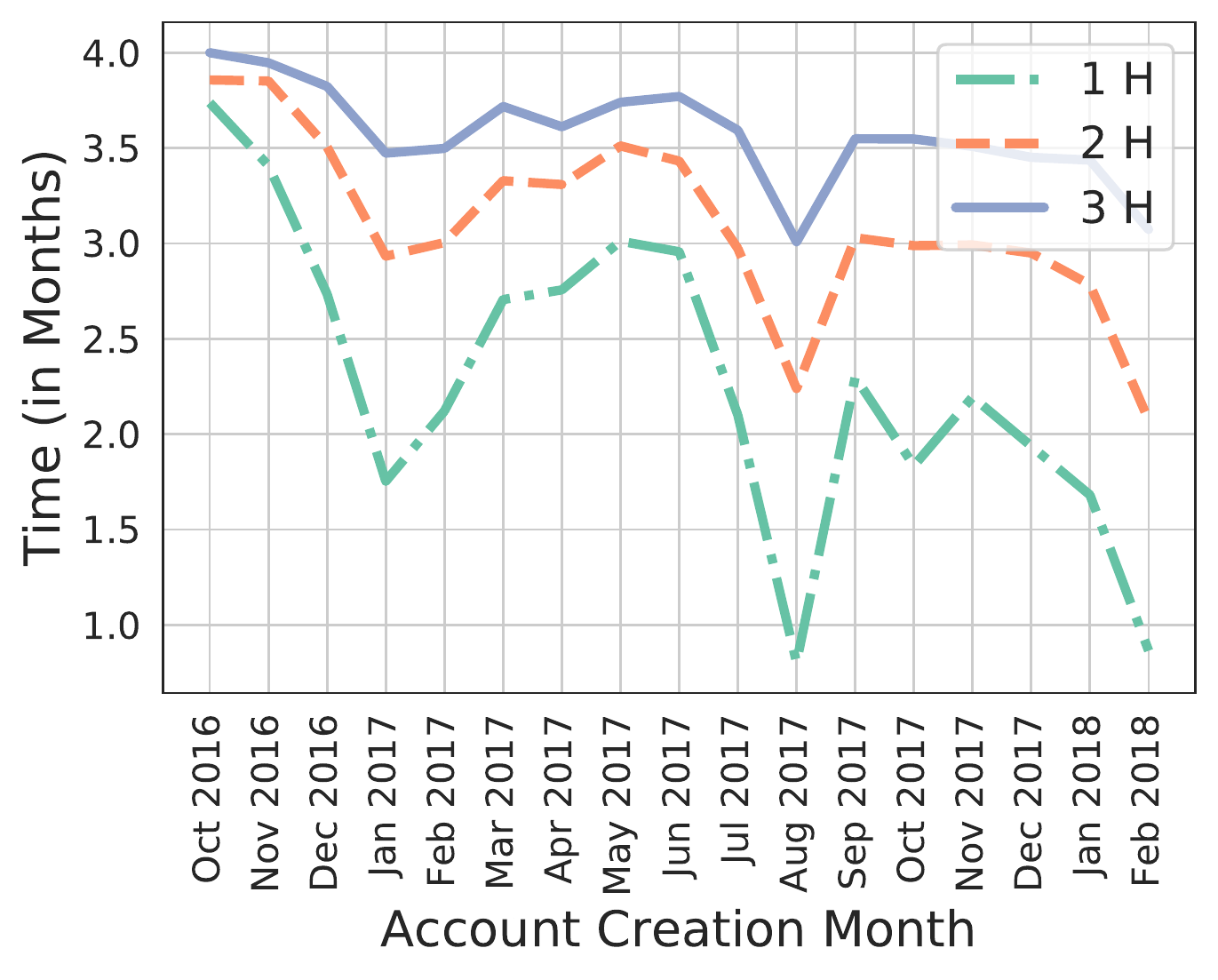}
		\caption{Threshold = 4}
	\end{subfigure}
	
	\begin{subfigure}[t]{0.45\textwidth}
        \includegraphics[width=\textwidth]{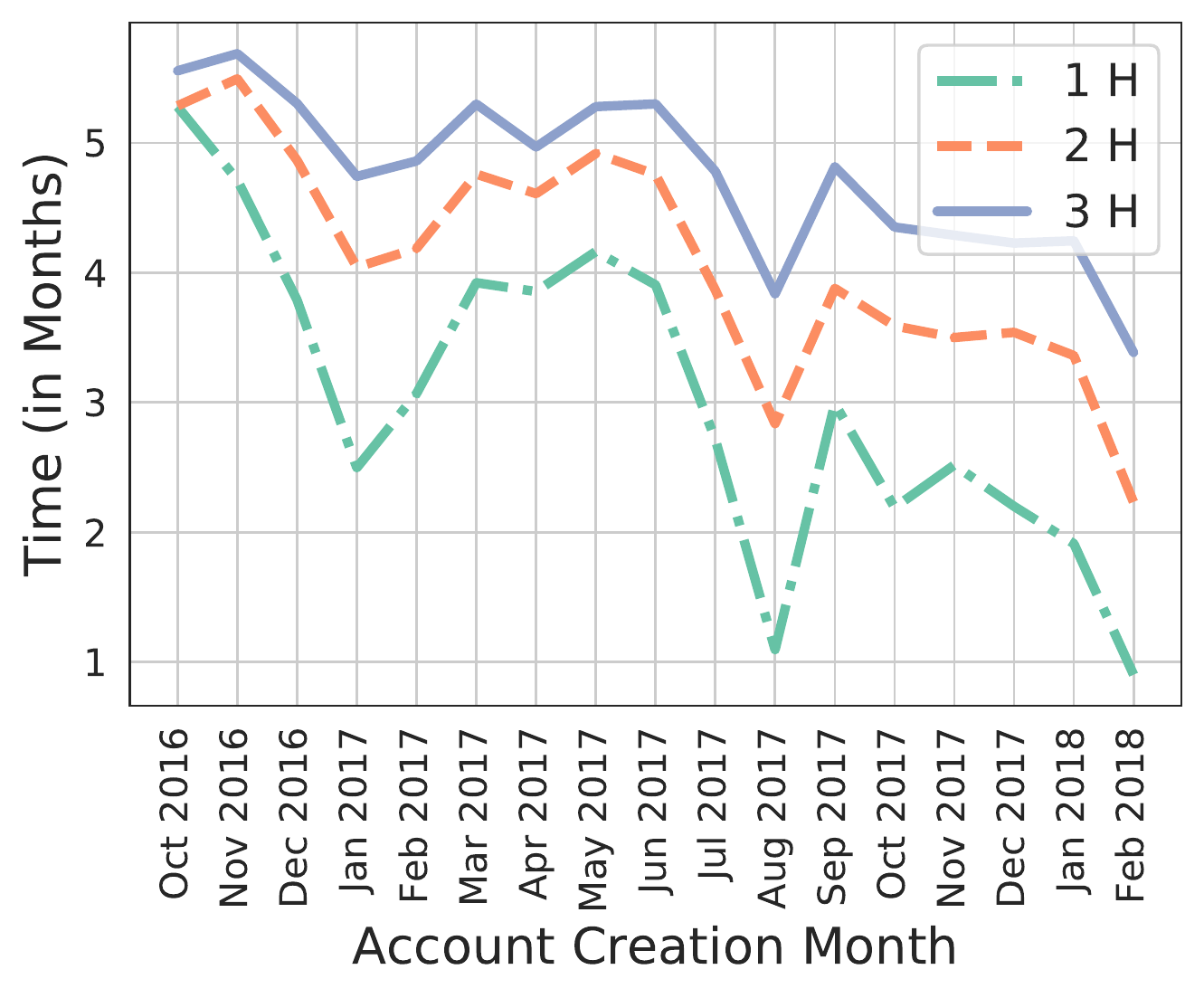}
	    \caption{Threshold = 5}
	\end{subfigure}
\begin{subfigure}[t]{0.45\textwidth}	        \includegraphics[width=\textwidth]{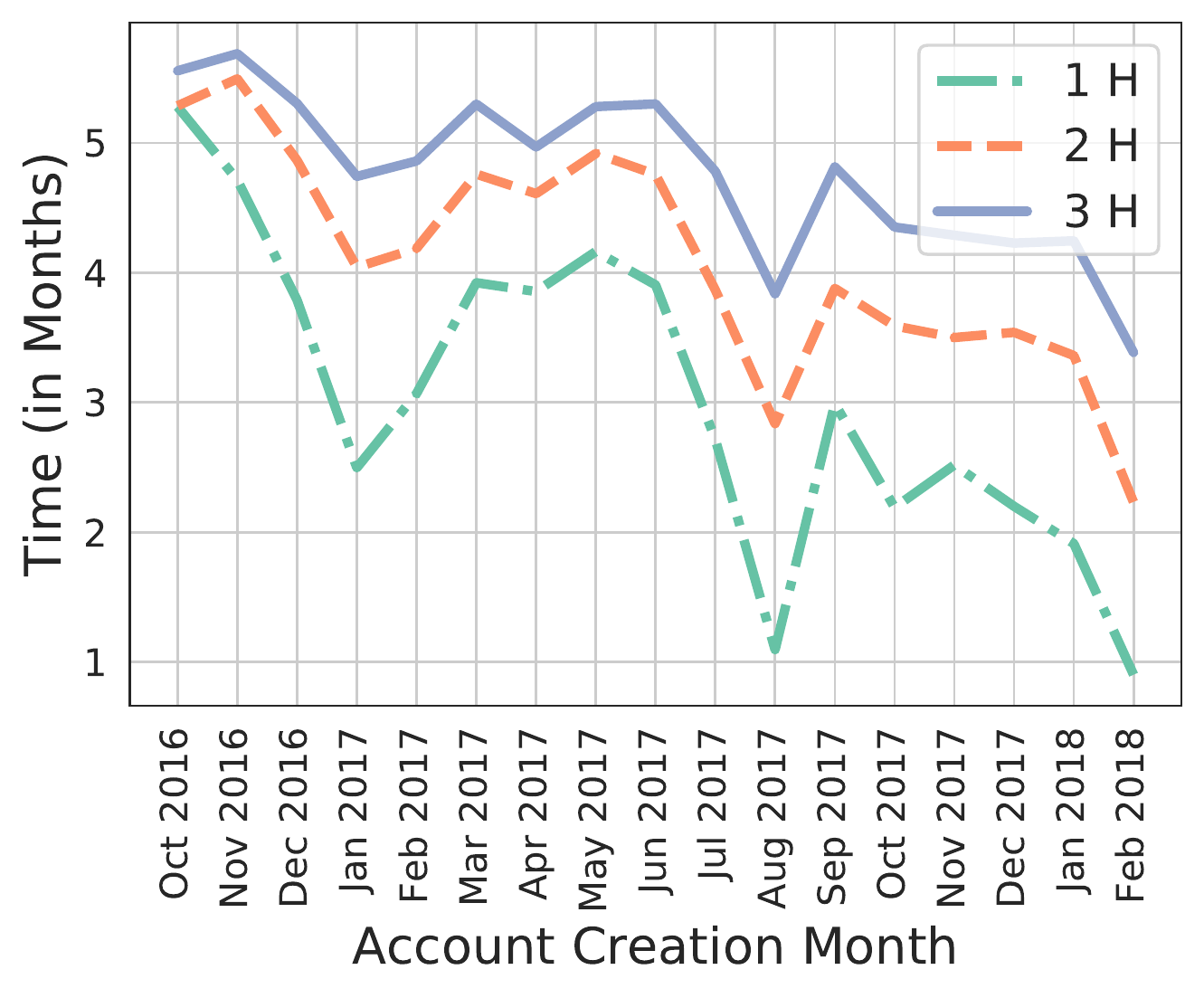}
		\caption{Threshold = 6}
		\label{fig:user_change_to_hate_time_6}
	\end{subfigure}
\caption{The figure shows the average amount of time (in months) that each user requires to achieve $\mathcal{N}$ `H' entries from his/her month of joining. The \textit{Threshold} refers to the upper limit for the number of months to achieve  $\mathcal{N}$ `H'. We can observe that as with time, the time required for a users to post his/her first hateful post is decreasing in Gab}
\Description{Figure with X-axis "Account Creation Month" and Y-axis "Time (in Months)"}
\label{fig:user_change_to_hate_time}
\end{figure*}

\subsection{New users are becoming hateful at a faster rate}
In Figure~\ref{fig:user_change_to_hate_time}, we show how much time does a user take to have the first, second and third `H' entry. We use a \textit{threshold} to ensure that the results are not biased as a user who joined in the initial months will have larger time to achieve the `H' entry as compared to users who joined later. 
Consider Figure~\ref{fig:user_change_to_hate_time_6} where we have used a value of six for \textit{threshold}. This means that a user who joins in March 2017, should achieve the first `H' within the next six months. If the user achieves the first `H' after that, (s)he would be considered to have achieved it in \textit{threshold} months (six in this case). Thus, if the users are achieving the first `H' after six months, the average value can be at max six.

We observe that with time, the time required for a user to become hateful decreases in Gab.

\section{RQ2: What was the impact of the hateful users on Gab?}

\subsection{Hate users receive replies much faster}

In this section we investigate the characteristics of the first replies obtained against a post.

\noindent\textbf{First reply time (FRT)}: The average time to the first reply to an originally hate post is 7.2 hours. In contrast, the average time to the first reply to an originally non-hate post is 10.4 hours.

\noindent\textbf{Distribution of first replies}: Out of the first replies to a hate post, 92\% of the replies are by a hateful user. In case of first replies to a non-hate post, non-hateful users cover 63\% of those replies.

\noindent\textbf{User level FRT}: Finally, in order to understand the user level engagement, we define FRT for a set of users $U$ as $ FRT_{U}= \frac{1}{|U|}\sum_{(u)\in U} T_u $, where $T_u$ represents the average time taken to get the first reply for the posts written by a user $u$.
We calculated the $FRT_U$ values for the set of hateful and non-hateful users and found that the average time for the first reply is 51.32 minutes for non-hate users, whereas it is 44.38 minutes for the hate users ($p$-value $<0.001$). This possibly indicates that the community is engaging with the hateful users at a faster speed as compared to the non-hateful users.

\subsection{Hateful users: lone wolf or clans}

In this section, we study the hateful and non-hateful users from a network-centric perspective by leveraging user-level dynamic graph.
This approach has been shown to be effective in extracting anomalous patterns in microblogging platforms such as Twitter and Weibo~\cite{zhang2015influenced,shin2016corescope}. In similar lines, we conduct a unique experiment, where we track the influence of hateful and non-hateful users across successive temporal snapshots.

\begin{figure*}[t]   	
	\centering
    
	\begin{subfigure}[t]{0.45\textwidth}
        \includegraphics[width=\textwidth]{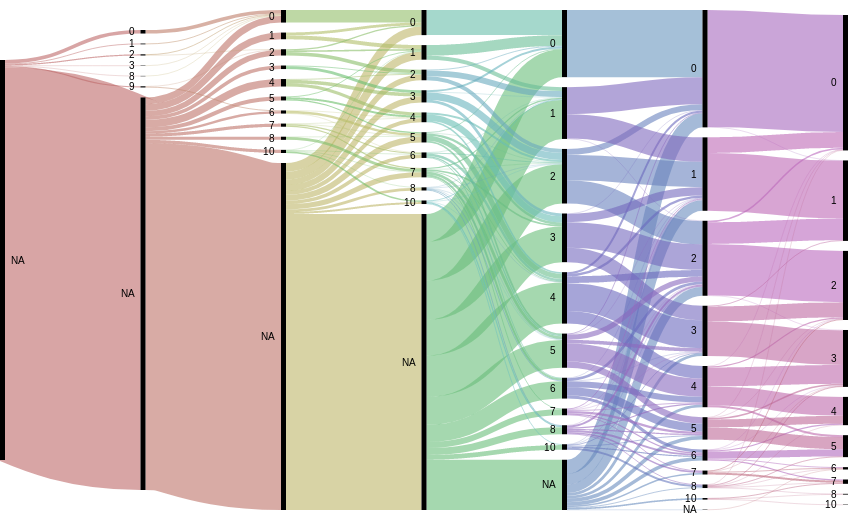}
	    \caption{Hateful user movement}
	    \label{fig:h-h}
	\end{subfigure}
\hspace{1cm}
\begin{subfigure}[t]{0.45\textwidth}	        \includegraphics[width=\textwidth]{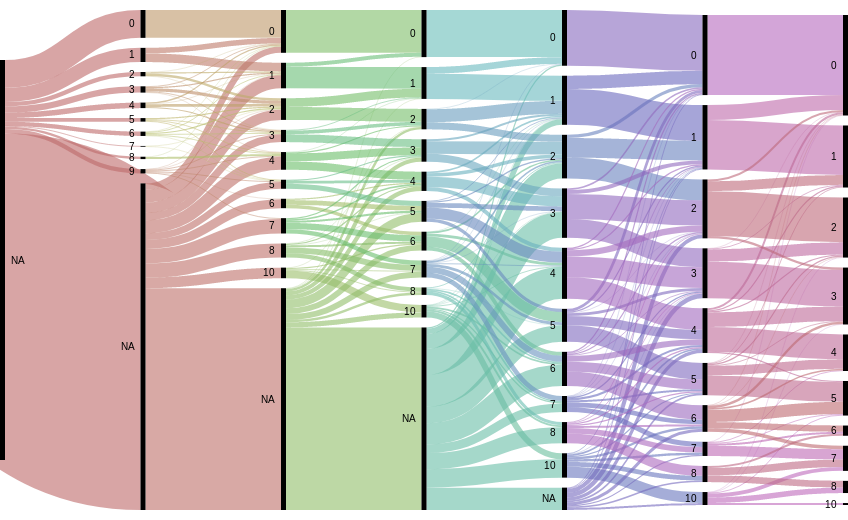}
		\caption{Non-hateful user movement}
		\label{fig:nh-nh}
	\end{subfigure}
\caption{Alluvial diagram to show the core-transition for the users. The stubs represent dynamic graph state after span of three months with the $1^\textrm{st}$ stub indicating $0^\textrm{th}$ month. A lower core value represents that a node is situated deeper in the network.}
\Description{Figure showing the alluvial diagram for the movement of hateful and non-hateful users. The values represent the core-periphery values.}
\label{Fig:Alluvial}
\end{figure*}

We utilize the node metric -- {\em k-core or coreness} to identify influential users in the network~\cite{shin2016corescope}. Nodes with high coreness are embedded in major information pathways. Hence they have been shown to be influential spreaders, that can diffuse information to a large portion of the network~\cite{malliaros2016locating,kitsak2010identification}. For further details about {\em coreness} and its several applications in functional role identification, we refer to Malliaros et al.~\cite{malliaros2019core}. We first calculate coreness of the undirected {\em follower/followee} graph for each temporal snapshot using {\em k-core decomposition}~\cite{malliaros2016locating}. In each snapshot, we subdivide all the nodes into $10$ buckets where consecutive buckets comprise increasing order of influential nodes, i.e., the bottom $10$ percentile nodes to the top $10$ percentile nodes in the network. We calculate the proportion of each category of users in all the proposed buckets across six temporal graphs. Each dynamic graph is an accumulation of users and connections (see Figure~\ref{fig:DynamicGraph}) over three months and aggregation is performed over a span of 18 months. We further estimate the proportion of migration from different buckets in consecutive snapshots. We illustrate the results as a flow diagram in Figure~\ref{Fig:Alluvial}. The innermost core is labeled 0, the next one labeled 1 and so on. The bars that have been annotated with a label $NA$ denote the proportion of users who have eventually been detected to be in a particular category but have not yet entered in the network at that time point (account is not yet created).

\noindent\textbf{Position of hateful users}: We demonstrate the flow of hateful users in Figure~\ref{fig:h-h}. The leftmost bar denotes the entire group strength. The following bars indicate consecutive time points, each showcasing the evolution of the network. 

We could observe several interesting patterns in Figure~\ref{fig:h-h}. In the initial three time points, we observe that a large proportion of users are confined to the outer shells of the network. This forms a network-centric validation of the hypothesis that newly joined users tend to familiarize themselves with the norms of the community and do not exert considerable influence~\cite{danescu2013no}. However, in the final time points we observe that the hateful users rapidly rise in ranks and the majority of them assimilate in the inner cores. This trend among Gab users has been found consistent with other microblogging sites like Twitter~\cite{ribeiro2018characterizing} where hate mongers have been found with higher eigenvector and betweenness centrality compared to normal accounts. There are also surprising cases where a fraction of users who have just joined the network, become part of the inner core very quickly. We believe that this is by their virtue of already knowing a lot of `inner core' Gab users even before they join the platform. 

\noindent\textbf{Position of non-hateful users}: Focusing on figure~\ref{fig:nh-nh}, which illustrates the case of non-hateful users, we see a contrasting trend. The flow diagram shows that users already in influential buckets continue to stay there over consecutive time periods. The increase in core size at a time point can be mostly attributed to the nodes of the nearby cores in the previous time point. We also observe that in the final snapshot of the graph outer cores (higher core number) tend to have a large fraction of users compared to hateful users (Figure~\ref{fig:h-h}).

\noindent\textbf{Acceleration toward the core}: We were also interested in understanding the rate at which the users were accelerating toward the core. To this end, we calculated the time it took for the users to reach bucket 0 from their account creation time. We found that a hateful user takes only 3.36 month on average to do this, whereas a non-hateful user requires 6.94 months on average to reach an inner core of the network. We tested the significance of this result with the Mann-Whitney $U$-test and found $U = 35203.5$ and $p$-value$=2.68e-06$.

\begin{figure}[!t]
\centering
\begin{minipage}[t]{.45\textwidth}
  \centering
  \includegraphics[width=.9\linewidth]{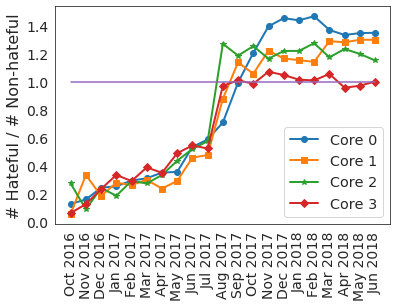}
  \captionof{figure}{The ratio of hateful users to non-hateful users for each month in four inner-most cores of the network. We observe that in the initial time periods, a higher proportion of non-hateful users were in the core part of the network, and in the later time period, the proportion of hateful users in the core became more.}
  \Description{Figure with X axis as months from October2016-June2018 and Y axis as the ratio of Hateful/Non-hateful users.}
\label{Fig:Core_transition}
\end{minipage}
\qquad
\begin{minipage}[t]{.45\textwidth}
  \centering
  \includegraphics[width=.9\linewidth]{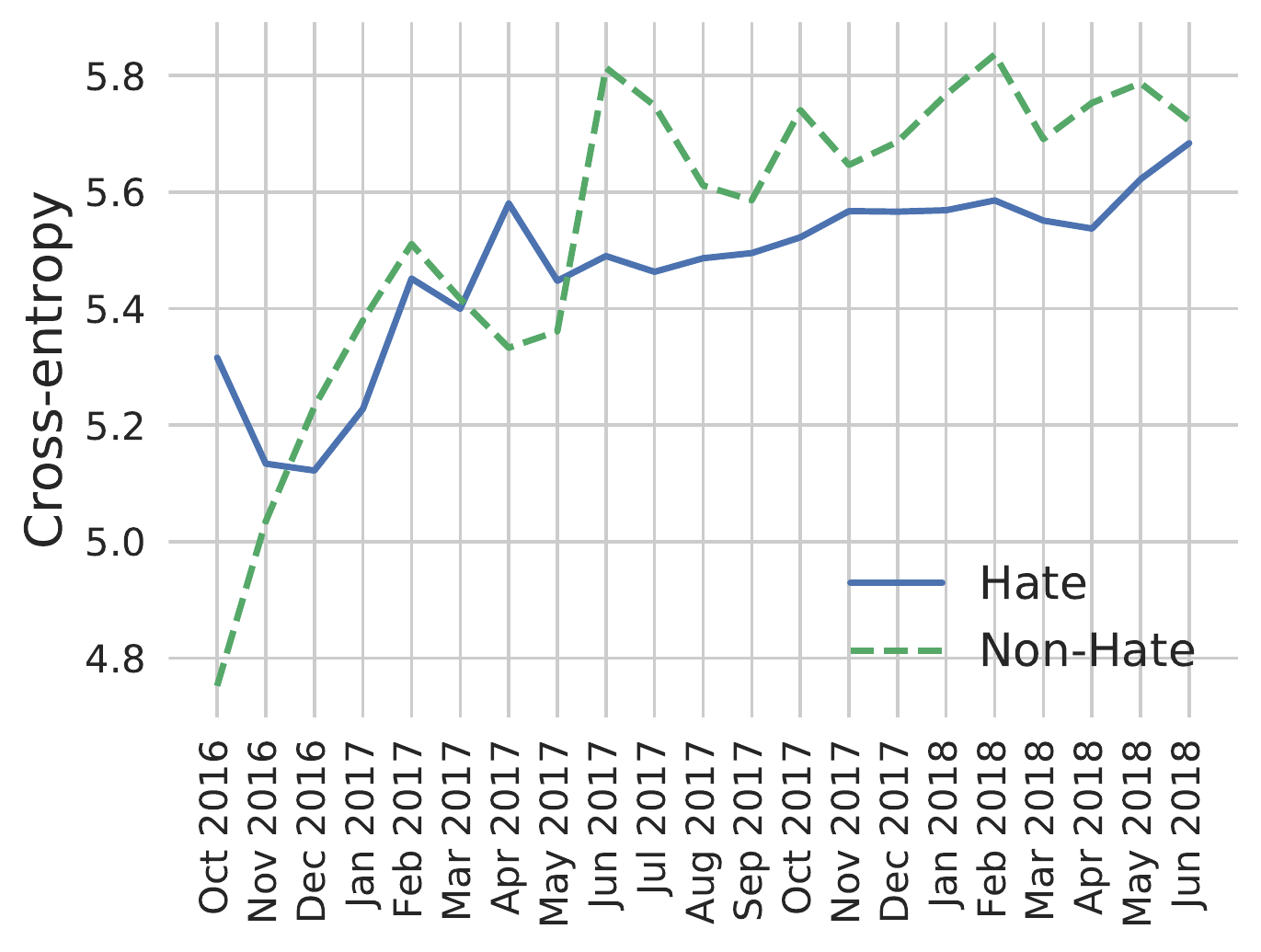}
  \captionof{figure}{Month wise cross-entropy of the predictions obtained from the H-SLM and N-SLM when the full data for that month is used as the test set. We observe that the language used by the Gab community became closer to the language used by the hateful users than the non-hateful ones.}
  \Description{Figure with X axis as months from October2016-June2018 and Y axis as the cross-entropy value of the hate and non-hate users. }
    \label{fig:entropy_month_wise}
\end{minipage}%
\end{figure}

To further understand the volume of users transitioning in-between the cores of the network, we compute the ratio of the hateful to the non-hateful users in a given core for each month. Figure~\ref{Fig:Core_transition} plots the ratio values. A value of 1.0 means that an equal number of hateful and non-hateful users occupy the same core in a particular month. A value less than one means that there were more non-hateful users in a particular core than there were hateful users. We observe that in the initial time periods (October 2016 - July 2017), the non-hateful users were occupying the inner core of the network more. However, after this, the fraction of hateful users in the innermost started increasing, and around August 2017 the fraction of hateful users surpassed the non-hateful ones. We observe similar trends in all the four innermost cores (0, 1, 2, and 3). The final trend following March 2018 (see Figure~\ref{Fig:Core_transition}) indicates that higher volume of hateful users occupy the inner cores of network, i.e., $Core_0 > Core_1 > Core_2 > Core_3$. Hence hateful users have higher propensity to occupy strategic positions in the network which facilitates information dissemination.

\subsection{Gab community is increasingly using language similar to users with high hate scores}

Gab started in August 2016 with the intent to become the `champion of free speech'. Since its inception, it has attracted several types of users. As the community evolves, so does the members in the community. To understand the temporal nature of the language of the users and the community, we utilize the framework developed by Danescu et al.~\cite{danescu2013no}. In their work, the authors use language models to track the linguistic change in communities.

We use kenLM~\cite{Heafield-estimate} to generate language models for each snapshot. These `Snapshot Language Models' (SLM) are generated for each month, and they capture the linguistic state of a community at one point of time. The SLMs allow us to capture how close a particular utterance is to a community. The `Hate Snapshot Language Model' (H-SLM) is generated using the posts written by the users with high hate score in a snapshot as the training data. Similarly, we generate the `Non-hate Snapshot Language Model' (N-SLM), which uses the posts written by users with low hate score in a snapshot for the training data. Note that unlike in the previous sections where we were building hate vectors aggregated over different time snapshots to call a user hateful/non-hateful, here we consider posts of users with high/low hate scores for a given snapshot to build the snapshot wise training data for the language models\footnote{It is not possible to extend the hate vector concept here as we are building language models snapshot by snapshot.}. For a given snapshot, we use the full data for testing. Using these two models, we test them on all the posts of the month and report the average cross entropy 
\[ H(d, \slm_{c_t}) = \frac{1}{|d|}\sum_{b_i \in d}\slm_ {c_t}(b_i) \]
where $H(.)$ represents the cross-entropy, $\slm_{c_t}(b_i)$ is the probability assigned to bigram $b_i$ from comment $d$ in community-month $c_t$. Here, the community can be hate (H-SLM) or non-hate (N-SLM)\footnote{We controlled for the spurious length effect by considering only the initial 30 words~\cite{danescu2013no}; the same controls are used in the cross-entropy calculations.}.

A higher value of cross-entropy indicates that the posts of the month deviate from the respective type of the community (hate/non-hate). We plot the entropy values in Figure~\ref{fig:entropy_month_wise}. As is observed, from around May 2017, the language used by the Gab community started getting closer to the language of users with high hate scores than the non-hateful users. This may indicate that the Gab community as a whole is having an increased correlation with the language used by the hateful users.

\subsection{Alternative matching criteria for selecting the control group of non-hateful users}

\begin{figure}
    \centering
    \includegraphics[width=0.5\linewidth]{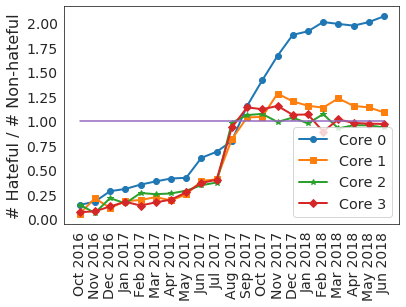}
    \caption{The ratio of hateful users to non-hateful users for each month in four inner-most cores of the network using the alternative matching criteria}
    \Description{Figure with X axis as months from October2016-June2018 and Y axis as the ratio of Hateful/Non-hateful users.}
    \label{fig:core-new-matching}
\end{figure}

To validate that our analysis were not biased because of a specific matching criteria used for the control group, we also considered an alternative matching criteria for selecting the non-hateful users by considering the distribution of activities, rather than just the mean\footnote{This is because mean might be influenced by the bursty nature of posts which is quite typical in various social media platforms.}. For each user, we first generate his/her posting distribution where each bin represents a month and the value in the bin represents the number of post/reposts done by the user in that month. We then calculate the mean and standard deviation of this distribution as the features for matching. For each hateful user, we find the closest non-hateful user using L1 distance between the corresponding feature vectors. Further, to ensure that the hateful and non-hateful users have similar distributions, we used M-W U test which gives a value of 0.497, indicating that the distributions are similar. Further, we re-ran the core-periphery results using the new matching pairs and report the results in Figure~\ref{fig:core-new-matching}. It shows that this new set also produces very similar results as reported in Figure \ref{Fig:Core_transition}. For the sake of brevity, we do not report other observations with this matching criteria which are very similar to the earlier ones.

\section{Discussion}

The debate surrounding `free speech' and `hate speech' has been around for several decades now. While free speech protects the right to express an opinion without censorship or restraint~\cite{howie2018protecting}, hate speech attacks a person or a group on the basis of attributes such as race, religion, ethnic origin, sexual orientation, disability, or gender~\cite{johnson2019hidden}. In the present work, we analyzed the temporal characteristics of hate speech on Gab.com, a site which allows freedom of speech and has been criticized for its lax moderation policy~\cite{zannettou2018gab}. We generate temporal snapshots of Gab and use DeGroot model to study the evolution of hate speech in Gab. In this section, we discuss the implications of our results for moderation, monitoring, and platform governance.

\subsection{Moderation}
Although the intent of Gab was to provide support for free speech, it is acting as a fertile ground for the fringe communities such as alt-right, neo-Nazis etc~\cite{zannettou2018gab}. We found many non-hateful accounts as well which share art and actively participate in discussions regarding politically sensitive topics such as transgender, immigration and other similar issues. These users could be affected by the hateful content posted in the community and thus our study is justified.
Moderation of speech is necessary at some level for the operation of online communication~\cite{langvardt2017regulating}. 
Article 19 of The International Covenant on Civil and Political Rights (ICCPR)~\cite{art19iccpr} protects the right to seek
and receive information of all kinds, regardless of frontiers, and through
any media. Companies like Twitter, Facebbok were called upon by United Nations' top experts for freedom of expression with the aim to align the company speech codes with the standard embodied in international human rights law, particularly ICCPR~\cite{aswad2018future}. However, it also gives State Parties the discretion to restrict expression if they can prove that each prong of a three-part test has been met. \citet{aswad2018future} studied the feasibility and desirability of aligning corporate speech codes with the ICCPR by focussing on Twitter's hate conduct policy. The article concludes with the observation that aligning the corporate speech codes with ICCPR outweigh the potential disadvantages.
Companies like Facebook have adopted methods such as counterspeech\footnote{\url{https://counterspeech.fb.com}} to combat the online hate speech\footnote{\url{https://techcrunch.com/2017/06/23/facebook-expands-its-hate-fighting-counterspeech-initiative-in-europe}}.
The basic idea is that instead of banning hate speech, we should use crowdsourced responses to reply to these messages~\cite{benesch2014countering, mathew2018thou,mathew2020interaction}. The main advantage of such an approach is that it does not violate the freedom of speech. However, there are some doubts on how much applicable/practical this approach is. Large scale studies would need to be done to observe the benefits and costs of such an approach.

We suggest that social media platforms could gamify an incentive mechanism~\cite{cavusoglu2015can,easley2016incentives} such as hierarchical badges (similar to stackoverflow) which are provided to users for counterspeech initiatives. 
They could also provide interface/tool to group moderators/users to identify hateful activities and take precautionary steps~\cite{mahar2018squadbox}. This would allow platforms to stop the spread of hateful messages in an early stage itself.

\subsection{Monitoring the growth of hate speech}
The platform should have interface which allows moderators to monitor the growth of hate speech in the community. This could be a crowdsourced effort which could help identify users who are attempting to spread hate speech.

As we have seen that the new users are gravitating toward the hateful community at a faster rate and quantity, there is a need for methods that could detect and prevent such movement. There could exist radicalization pipelines~\cite{ribeiro2019auditing} which could navigate a user toward hateful contents. Platforms should make sure that their user feed and recommendation algorithms are free from such issues. Exposure to such content could also lead to desensitization toward the victim community~\cite{soral2018exposure}. We would need methods which would take the user network into consideration as well. Instead of waiting for a user to post his/her hateful post after the indoctrination, the platforms will need to be proactive instead of reactive. Some simple methods such as nudge~\cite{thaler2009nudge}, or changing the user feed to reduce polarization~\cite{celis2019controlling} could be an initial step. Further research is required in this area to study these points more carefully.

\subsection{Applicability on other social media platforms}

While our methods might not work directly on other platforms such as Twitter and Facebook, it could act as a good starting point for such analysis. However the more important question is that if we were able to detect the presence of hate speech in these other platforms what would be the most effective strategy to combat with it? Can our results shed some light on how the flow of hate speech could have been impeded? While the current design recommendations on these platforms is a flat suspension of the account, can our results suggest less harsh but more effective regulations?  

The first thing that our study points out is that early signals feature in the text as well as the network of the users. So, the intervention process also can be started early and any specific platform can take appropriate steps to shield these hate messages from the rest of the network. Similarly, any platform needs to monitor the new set of users joining; since many of these users are potentially vulnerable\footnote{Although there might be a fraction of users who join already with an ingrained hate ideology.}, policies should be made to keep them far from the `hate core'. Finally since language is one of the biggest ammunition of any community, any platform needs to constantly graduate the `temper' of the language used by the users and ensure that it does not violate the decorum of online speech. A regular feedback system can be set up to extensively promote healthy and abuse-free discussion among the users.

\subsection{Platform governance -- the rising role of CSCW}
All the points that we had discussed above related to moderation and monitoring can be aptly summarized as initiatives toward platform governance. We believe that within this initiative the CSCW design principles of the social media platforms need to be completely overhauled. In February 2019, the United Kingdom's Digital, Media, Culture, and Sport (DCMS) committee issued a verdict that social media platforms can no longer hide themselves behind the claim that they are merely a `platform' and therefore have no responsibility of regulating the content of their sites\footnote{https://policyreview.info/articles/analysis/platform-governance-triangle-conceptualising-informal-regulation-online-content}. In fact, the European Union now has the `EU Code of Conduct on Terror and Hate Content' (CoT) that applies to the entire EU region. Despite the increase in toxic content and harassment, Twitter did not have a policy of their own to mitigate these issues until the company created a new organisation -- `Twitter Trust and Safety Council' in 2015. A common way deployed by the EU to combat such online hate content involves creation of working groups that combine voices from different avenues including academia, industry and civil society. For instance, in January 2018, 39 experts met to frame the `Code of Practice on Online Disinformation' which was signed by tech giants like Facebook, Google etc. We believe that CSCW practitioners have a lead role to play in such committees and any code of conduct cannot materialize unless the CSCW design policies of these platforms are reexamined from scratch.

\subsection{Events around August 2017}

We can observe from several results (Figure 7, 9, 10, and 12) that there are sudden variations around August 2017. We attempted to investigate the possible reasons for this. Often hate speech may be an aftermath of some real-world events~\cite{Olteanu2018TheEO} and hashtags could be an important indicator of such events. For the month of August 2017, we extract all the hashtags used along with their frequency. We tried to analyse the hashtags that were more prominent among hate users as compared to non-hate users. To do this we rank the hashtags in the posts of hate users based on their frequency, which we call \textit{Rank-Hate}. Similarly, we created a ranked list \textit{Rank-NonHate}, for hashtags in the post of non-hate users. For each hashtag a final score is calculated based on its positional difference in the \textit{Rank-NonHate} and \textit{Rank-Hate}. The more positive this score the more prominent the hashtag is in the hateful users' posts. In this way the top 5 hashtags were  \#Charlottesville, \#NewRight, \#Antifa, \#UniteTheRight and \#AltLeft. On examining the posts using these hashtags, we found that they are linked to \textit{Unite the Right Rally}\footnote{\url{https://en.wikipedia.org/wiki/Unite_the_Right_rally}} which was a white supremacist and neo-Nazi\footnote{\url{www.vox.com/2017/8/12/16138246/charlottesville-nazi-rally-right-uva}} rally that was conducted in Charlottesville, Virginia, from August 11 to 12, 2017. The event lead to several users joining Gab and using hateful language (Figure 9 and 10).

\subsection{Effects of Twitter purge on December 2017}
We can observe from Figure~\ref{fig:hate_distribution} that there was a sudden rise in the mid and high hate users accounts in Gab. Incidentally, this coincided with the Twitter purging several prominent accounts\footnote{\url{www.vox.com/2017/12/18/16790864/twitter-bans-nazis-hate-groups}}. We first looked into the users who joined Gab in December 2017, and were assigned mid and high hate score. We found that only 435 (0.63\%) out of 69,236 mid hate users of Dec 2017 had joined Gab in Dec 2017 itself. Similarly, only 262 (2.31\%) out of 11,326  high hate users of Dec 2017 had joined in Dec 2017 itself. This means that a very small fraction of users belonged to the new user accounts and majority of the users that were categorized as mid and high hate were already existing users of Gab. To further verify this, we re-run the core-periphery experiments where we look into the ratio of hateful users to non-hateful users for each month in a specific core of the network.  In this experiment we do not consider the users who joined Gab in Dec 2017. We observe that removing these users had little to no effect on the results. Further, it is also possible that the new users could have joined the Gab network later in January, 2018. To check this we repeat the above experiment and remove the newly joined users in December 2017 and January 2018. Again, we get very similar results implying that the results were not affected by the Twitter purge activity.

\subsection{Ethical considerations and implications}

The discussion surrounding whether hate speech should banned or protected under freedom of speech is of great significance~\cite{howard2019free,gelber2002speaking,newman2017finding,delgado2018must}. This issue is further complicated by the varying interpretation of what constitutes a hate speech~\cite{brown2017hate}. Online platforms such as Twitter\footnote{\url{https://help.twitter.com/en/rules-and-policies/hateful-conduct-policy}} and Facebook\footnote{\url{https://www.facebook.com/communitystandards/hate_speech}} have defined rules which act as guidelines to decide if a post is a hate speech. Whereas on Gab, such guidelines do not exist. In this work, we provide a peek into the hate ecosystem developed on a platform where the main moderation is `self-censorship'.

We caution against our work being perceived as a means for full-scale censorship. Our work is not indented to be perceived as a support for full-scale censorship. We simply argue that the `free flow' of hate speech may be monitored. We leave it to the platform or government to implement a system that would reduce such speech in the online space.

\section{Limitations and future works}

There are several limitations of our work. We are well aware that studies conducted on only one social media such as Gab have certain limitations and drawbacks. Especially, since other social media sites delete/suspend hateful posts and/or users, it becomes hard to conduct similar studies on those platforms. The initial keywords selected for the hate users were in English. This would bias the initial belief value assignment as users who use non-English hate speech would not be detected directly. But, since these users follow similar hate users and repost several of their content, we would still detect many of them. We plan to take up the multilingual aspect as an immediate future work. Another major limitation of our work is the high-precision focus of the work which would leave out several users who could have been hateful.

As part of the future work, we also plan to use the discourse structure of these hateful users for better understanding of the tactics used by users in spreading hate speech~\cite{phadke2018framing}. This would allow us to break down the hate speech discourse into multiple components and study them in detail.

\section{Conclusion}
In this paper, we perform the first temporal analysis of hate speech in online social media. Using an extensive dataset of 21M posts by 314K users on Gab, we divide the dataset into multiple snapshots and assign a hate score to each user at every snapshot. We then check for variations in the hate score of the user. We characterize these account types on the basis of text and network structure. We observe that a large fraction of hateful users occupy the core of the Gab network and they reach the core at a much faster rate as compared to non-hateful users. The language of the community as a whole is getting more correlated to that of the hateful users as compared to the non-hateful users. Our work would be useful to platform designers to detect the hateful users at an early stage and introduce appropriate measures to moderate the users' stance.

\bibliographystyle{ACM-Reference-Format}
\bibliography{main}


\end{document}